\documentclass[12pt]{article}
\usepackage{amsmath,amssymb,amsfonts,amscd,bbm,graphicx,cite}
\newcommand{\ie}{{\it i.e.}}

\newcommand{\eg}{{\it e.g.}}

\newcommand{\cf}{{\it cf.}}

\newcommand{\eq}{Eq.}
\newcommand{\eqs}{Eqs.}

\newcommand{\Fig}{Fig.}

\newcommand{\Figs}{Figs.}
\newcommand{\Ref}{Ref.}
\newcommand{\Refs}{Refs.}
\newcommand{\Sec}{Sec.}

\newcommand{\Tab}{Table}

\begin{document} 


\thispagestyle{empty}
\renewcommand{\thefootnote}{\fnsymbol{footnote}}
\setcounter{footnote}{1}

\vspace*{-1.cm}
\begin{flushright}
OSU-HEP-04-13
\end{flushright}
\vspace*{1.8cm}

\centerline{\Large\bf Neutrino Oscillations in}
\vspace*{3mm}
\centerline{\Large\bf Deconstructed Dimensions}

\vspace*{18mm}

\centerline{\large\bf 
Tomas H\"allgren$^a$\footnote{E-mail: \texttt{tomash@theophys.kth.se}},
Tommy Ohlsson$^a$\footnote{E-mail: \texttt{tommy@theophys.kth.se}},
and 
Gerhart Seidl$^b$\footnote{E-mail: \texttt{gseidl@susygut.phy.okstate.edu}}}

\vspace*{5mm}
\begin{center}
$^a${\em Division of Mathematical Physics, Department of Physics}\\
{\em Royal Institute of Technology (KTH) -- AlbaNova University Center}\\
{\em Roslagstullsbacken 11, 106~91 Stockholm, Sweden}\\
~\\
$^b${\em Department of Physics, Oklahoma State University}\\
{\em Stillwater, OK 74078, USA}
\end{center}

\vspace*{20mm}

\centerline{\bf Abstract}
\vspace*{2mm}
We present a model for neutrino oscillations in the
presence of a deconstructed non-gravitational large extra dimension
compactified on the boundary of a two-dimensional disk. In the deconstructed phase, sub-mm lattice
spacings are generated from the hierarchy of energy scales between
$\sim 1\:{\rm TeV}$ and the usual $B-L$ breaking scale
$\sim 10^{15}\:{\rm GeV}$. Here, small short-distance cutoffs down to
$\sim 1\:{\rm eV}$ can be motivated
by the strong coupling behavior of gravity in local discrete extra
dimensions. This could make it possible to probe the discretization of
extra dimensions and non-trivial field configurations in theory spaces
which have only a few sites, \ie, for coarse
latticizations. Thus, the model has relevance to present and future
precision neutrino oscillation experiments.

\renewcommand{\thefootnote}{\arabic{footnote}}
\setcounter{footnote}{0}

\newpage

\section{Introduction}\label{sec:Introduction}

There may be many ways to give a sensible short-distance definition of
a model, which describes the Kaluza--Klein (KK) modes
\cite{Kaluza:1921tu} of a compactified higher-dimensional gauge
theory. One attractive possibility is offered by deconstructed
\cite{Arkani-Hamed:2001ca} or latticized \cite{Hill:2000mu} extra
dimensions. Deconstruction provides a new type of four-dimensional
(4D) manifestly gauge-invariant and renormalizable field
theories\footnote{For an early formulation in terms of infinite arrays
of gauge theories, see \Ref~\cite{Halpern:1975yj}.}, which simulates
in the infrared (IR) the physics of extra dimensions, and thus, yields a
possible ultraviolet (UV) completion of higher-dimensional gauge
theories \cite{UVcompletion}. In deconstruction, a bulk gauge
symmetry is mapped onto $N$ copies of a 4D gauge group with
bi-fundamental Higgs ``link'' fields connecting the neighboring
groups. This becomes equivalent with treating the extra dimensional
space as a transverse lattice \cite{Bardeen:1976tm}, where the inverse
lattice spacing $\Lambda$ is set by the vacuum expectation values
(VEV's) of the link fields. These models find a graphical
interpretation in ``theory space'' \cite{Arkani-Hamed:2001ed}, where
the notion of extra dimensions has been translated into more general
organizing principles of 4D field theories \cite{Hill:2002rb}. With
this emphasis, the idea of theory space has stimulated the development
of interesting theories in four dimensions, which need not be
attributed any direct extra-dimensional correspondence at all.

Since deconstruction is formulated within conventional 4D field
theory, one may now expect that a typical lattice cutoff $\Lambda$
naturally varies in the UV desert between $\sim 1\:{\rm TeV}$ and
$M_{Pl}\simeq 10^{18}\:{\rm GeV}$, \ie, in the energy range where
usual effective field theories are defined. Indeed, if $\Lambda$ is,
\eg, of the order the $B-L$ breaking scale $M_{B-L}\simeq
10^{15}\:{\rm GeV}$, then a realistic neutrino phenomenology can
arise \cite{Balaji:2003st}, where small neutrino masses and bilarge leptonic mixing result
from the lattice-link structure of the theory itself.\footnote{A recent attempt to obtain the quark mixing from deconstruction has been described
in Ref.~\cite{Hung:2004wf}.} Generally, deconstructed non-gravitational extra
dimensions always give a sensible effective field theory up to energy
scales of the order $4\pi \Lambda$, and hence, it seems that low
cutoff scales $\Lambda\ll 1\:{\rm TeV}$ are not particularly preferred
in this case. However, for inverse lattice spacings above a TeV it
will be difficult to test theory space models at low energies, unless
the number of lattice sites $N$ is very large. On the other hand,
cutoff scales which are hierarchically small compared to $\sim 1\:{\rm TeV}$
emerge in local theory space formulations of
gravity. Local discrete gravitational extra dimensions are
characterized by an intrinsic maximal inverse lattice spacing
$\Lambda_{\rm max}$, which is related by a ``UV/IR connection'' to the
higher-dimensional Planck scale $M_\ast$ and the compactification
scale $R^{-1}$
\cite{Arkani-Hamed:2002sp,Arkani-Hamed:2003vb,Schwartz:2003vj}. Owing
to this UV/IR connection, one obtains a small cutoff $\Lambda_{\rm
max}\ll M_\ast$, when $R^{-1}$ and $M_\ast$ are clearly separated. Low
string scale models with $M_\ast\sim 1\:{\rm TeV}$, like the large
extra dimensional scenario of Arkani-Hamed, Dimopoulos, and Dvali
(ADD) \cite{Arkani-Hamed:1998rs} (see, alternatively,
Refs.~\cite{Randall:1999ee, CremadesKokorelisx}), will thus lead to a hierarchy
$\Lambda_{\rm max}\ll 1\:{\rm TeV}$, when formulated in local theory
spaces.

This could, in principle, open up the possibility to test these models
experimentally already at low energies. 
For instance, it is well known that sub-mm
sized extra dimensions would allow for a measurable conversion of the active
neutrinos in higher-dimensional neutrino oscillations
\cite{Arkani-Hamed:1998vp,Dienes:1998sb,Dvali:1999cn,Mohapatra:1999zd,Mohapatra:2001,MoreauGrx:2004}. Therefore, if discrete gravitational or non-gravitational large extra
dimensions exhibit a cutoff $\Lambda_{\rm max}\sim R^{-1}\sim ({\rm mm})^{-1}$, one could study with neutrino oscillations the finite
discretization effects of few site models far away from the continuum
limit. Clearly, in any such model, it would be important to have a
dynamical understanding of the smallness of the inverse lattice
spacing $\Lambda\lesssim \Lambda_{\rm max}$ in terms of a mechanism,
which generates a value $\Lambda\sim (\mu{\rm m})^{-1}$ from energy
scales in the UV desert above a TeV.

In this paper, we use neutrino oscillations to probe deconstructed
non-gravitational large extra dimensions, which have inverse lattice
spacings of the order $(\mu{\rm m})^{-1}$. We motivate the relevance
of small inverse lattice spacings $\ll 1\text{TeV}$ by the strong-coupling behavior of gravity in a
six-dimensional (6D) model, where the two extra dimensions have been
naively discretized in a local theory space. In fact, by requiring that a model of this
type with only a few sites allows for testable predictions in typical
Cavendish-like (laboratory) experiments, the maximal strong coupling scale
$\Lambda_{\rm max}$ is found to be $\Lambda_\text{\rm max}\lesssim 1\,\text{keV}$. We analyze a toy model with few sites for a deconstructed
$U(1)$ gauge theory on a disk, where a sub-mm sized boundary emerges
dynamically from the hierarchy of energy scales between $\sim 1\:{\rm TeV}$
and $M_{B-L}\simeq 10^{15}\:{\rm GeV}$. The active
Standard Model
(SM) neutrinos mix with a latticized right-handed (\ie, SM
singlet) neutrino, which propagates on the boundary of the disk. A
gaugeable cyclic discrete symmetry ensures that the latticized
neutrino can be treated as a massless Wilson fermion.  This symmetry
also establishes a Wilson-line type effective coupling between the
active neutrinos and the latticized right-handed neutrino. Hence, the
model reproduces for coarse latticizations major features of the 5D ADD
continuum theory with one right-handed neutrino in
the bulk, which couples to the SM through a local interaction. We study the possible neutrino oscillation patterns for the
cases of a twisted/untwisted right-handed neutrino and an even/odd
number of lattice sites. The neutrino oscillations may account for
possible subdominant deviations of neutrino oscillations from the
solution to the solar and atmospheric neutrino anomalies
and could be tested in ongoing and future low-energy neutrino
experiments.

The paper is organized as follows. In \Sec~\ref{sec:gravity}, we
review the strong coupling behavior of gravity in local theory spaces
for the example two discretized gravitational extra dimensions. Next,
in \Sec~\ref{sec:Model}, we present our model for the deconstruction
of a $U(1)$ gauge theory on the boundary of a two-dimensional
disk. This model generates from the hierarchy of energy scales between
$\sim 1\:{\rm TeV}$ and $M_{B-L}$ lattice spacings in the
sub-mm range, and thus, represents a few site model for large extra
dimensions. Properties of the mass spectrum and the coupling of a
latticized right-handed neutrino which propagates on the boundary of the disk
are analyzed in \Sec~\ref{sec:neutrinomasses}. The mixing of the KK modes of
the latticized neutrino with the active neutrinos is calculated in
\Sec~\ref{sec:bulkmodes}. In \Sec~\ref{sec:neutrinooscillations}, we
consider the neutrino oscillations of the active neutrinos into the
latticized fermion and discuss the different neutrino oscillation
patterns. Finally, in \Sec~\ref{sec:disc}, we present our summary and conclusions. In Appendix \ref{app:anomalies}, we study the cancellation of
anomalies in our model, and in Appendix \ref{app:diagonalization}, we
describe in some more detail the neutrino mixing matrices.

\section{Strong coupling in discretized 6D gravity}\label{sec:gravity}

In this section, we will estimate the strong coupling scale in a naive
discretization of 6D gravity. It turns out that in local theory spaces
with sub-mm sized extra dimensions, the short-distance cutoff, as
determined from graviton scattering, lies in the sub-MeV-range, which is far away from the (effective) 6D Planck scale $M_{6D}\sim
1\:{\rm TeV}$. If we require
that such a model leads to predictions which are testable in
Cavendish-like experiments, then the UV cutoff can be even further lowered
by several orders of magnitude. This motivates to consider the
extreme limit of lattice cutoffs in the eV-range. This section briefly reviews part of
the formulation of gravity in theory space as given in
\Ref~\cite{Arkani-Hamed:2002sp} by considering the case of two
discrete gravitational extra dimensions. In doing so, we follow
closely the treatment of a single discrete gravitational extra
dimension in \Refs~\cite{Arkani-Hamed:2003vb,Schwartz:2003vj}.

\begin{figure}
\begin{center}
\includegraphics*[bb = 235 575 375 715,height=4.0cm]{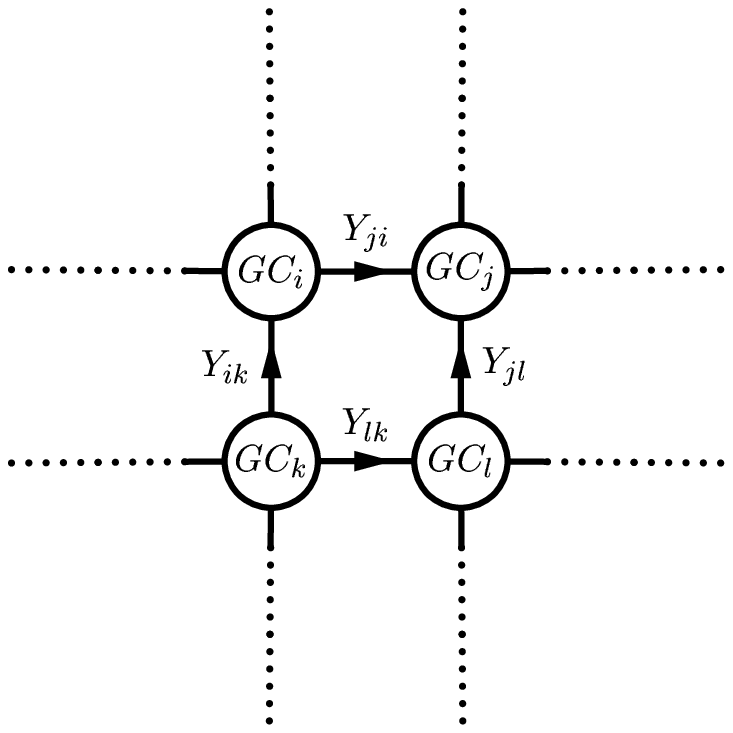}
\end{center}
\vspace*{-5mm}
\caption{\small{Plaquette in the theory space for a naive
discretization of 6D gravity. Each circle denotes one general
coordinate invariance ($GC$). Two neighboring GC's $GC_a$ and $GC_b$,
where $|b-a|=1$, are connected by a link field $Y_{ba}$, which can be
regarded as a map from site $a$ to site $b$.}}\label{fig:grid}
\end{figure}

The starting point of our discussion is standard general relativity in
six dimensions, where the two extra spatial dimensions have been
compactified on a square. We write the coordinates in the 6D space as
$z_M=(x_\mu,y_k)$, where the 6D Lorentz indices are denoted by upper
case Roman letters $M=0,1,2,3,5,6$, while we use for the usual 4D
Lorentz indices lower case Greek letters $\mu=0,1,2,3$, and the
coordinates $y_k$ $(k=1,2)$ describe the fifth and sixth dimension.
We write the 6D metric $G_{MN}(x_\mu,y_k)$ in block form as
\begin{equation}\label{eq:metric}
 G_{MN}(x_\mu,y_k)=
\left(
\begin{matrix}
g_{\mu\nu}(x_\mu,y_k)&0\\
0&\mathbbm{1}_2
\end{matrix}\right),
\end{equation}
where we have neglected the associated spin-1 and spin-0
excitations\footnote{The spin-1 states decouple from the fields
confined to the 3-brane, while the spin-0 fields interact only through
the dilaton mode \cite{Han:1998sg}.}. For our purposes, it is
sufficient to restrict $G_{MN}$ to the simplified form shown in
\eq~(\ref{eq:metric}), since we will be only concerned here with
the leading order UV behavior of the scattering amplitudes in the
effective theory for massive gravitons.  The 6D Einstein--Hilbert
action $\mathcal{S}_{\rm EH}=\int d^6x\:M_{6D}^4\sqrt{|G|}R_{6D}[G]$,
in which $R_{nD}[G]$ is the Ricci scalar in $n$ dimensions, can then be
rewritten as
\begin{equation}\label{eq:EH}
 \mathcal{S}_{\rm EH}=\int d^4x\:dy_1\:dy_2\sqrt{|g|}M_{6D}^4(
R_{4D}[g]+\frac{1}{4}\partial_{y_k}g_{\mu\nu}
(g^{\mu\alpha}g^{\nu\beta}-g^{\mu\nu}g^{\alpha\beta})\partial_{y_k}g_{\alpha\beta}),
\end{equation} 
where summation over $k=1,2$ is understood. In order to simulate the effects of
the two extra dimensions in a 4D model, we assume $N^2$ copies of 4D
general coordinate invariance (GC), which we denote as $GC_i$, where
$i=(i_1,i_2)$ and $i_1,i_2=1,2,\ldots,N$. In theory space, each
coordinate invariance $GC_i$ is represented by one ``site'' $i$, where
any two neighboring sites $GC_i$ and $GC_j$ with $|j-i|=1$ are
connected by a link field $Y_{ji}$. Thus, the collection of sites and
links forms a two-dimensional transverse lattice (see
\Fig~\ref{fig:grid}).
As already pointed out in \Ref~\cite{Hill:2000mu}, this can be simply
interpreted as a variation of the Eguchi-Kawai plaquette model for
large $N$ gauge theories \cite{Eguchi:1982nm}.

Each site $i$ is equipped with its own metric $g_{\mu\nu}^i(x_i)$,
which is a function of the 4D coordinates $x_i\equiv (x_i^\mu)$ for
this site. In order to obtain a lattice version of the derivatives
$\partial_{y_k}$ in \eq~(\ref{eq:EH}), we will employ
appropriate pullbacks of the coordinates $x^\mu_i$ and metrics
$g_{\mu\nu}^i$ as defined in \Ref~\cite{Arkani-Hamed:2002sp}. In this
line of thought, a link field $Y_{ji}$ can be considered as a pullback
function $Y_{ji}^\mu(x_i)$, which maps a vector $x_i$ on the site $i$
onto a vector $x_j$ on the site $j$ with coordinates
$x_j^\mu=Y^\mu_{ji}(x_i)$. Likewise, to compare the metric
$g^j_{\mu\nu}(x_j)$ on the site $j$ with the metric
$g^i_{\mu\nu}(x_i)$ on the site $i$, one can introduce the field
$G_{\mu\nu}^{ji}(x_i)\equiv(\partial Y^\alpha_{ji}/\partial x^\mu_i)
(\partial Y^\beta_{ji}/\partial
x^\nu_i)g_{\alpha\beta}^{j}(Y_{ji}(x_i)) $, which transforms as a
metric tensor under $GC_i$, but is left fixed by $GC_j$. Therefore,
$G^{ji}$ is the pullback of the metric $g^j_{\mu\nu}(x_j)$ on the site
$j$ to the site $i$.

Assuming equal lattice spacings $a$ in both the $y_1$- and
$y_2$-directions, we can now associate the derivatives in the extra
dimensional continuum theory with the nearest neighbor forward
difference operators $\partial g_{\mu\nu}/\partial y_k=a^{-1}\left[
g^i_{\mu\nu}(x_i)-G^{ji}_{\mu\nu}(x_i)\right]$, where
$(j_1,j_2)=(i_1+1,i_2)$ for $k=1$ and $(j_1,j_2)=(i_1,i_2+1)$ for
$k=2$. With this definition, we obtain from \eq~(\ref{eq:EH}) a naive
transverse lattice formulation of the 6D Einstein--Hilbert action
\begin{eqnarray}\label{eq:discretized}
 \mathcal{S}&=&
\sum_{i,j}\int d^4x_i\sqrt{|g^i|}M^2\left(
R_{4D}[g^i]+a^{-2}
(g_{\mu\nu}^i(x_i)-G^{ji}_{\mu\nu}(x_i))\right.\nonumber\\
&&\times\left.
(g_{\alpha\beta}^i(x_i)-G^{ji}_{\alpha\beta}(x_i))
(g^{i\mu\nu}g^{i\alpha\beta}-g^{i\mu\alpha}g^{i\nu\beta})
\right),
\end{eqnarray}
where $|j-i|=1$ and $M$ is the fundamental 4D Planck scale, which can,
in general, be different from the usual 4D Planck scale $M_{Pl}\simeq
10^{18}\:{\rm GeV}$. Comparison with \eq~(\ref{eq:EH}) shows that
$R=Na$, where $\sim 1/R$ is the compactification scale,
$M_{6D}^4=M^2a^{-2}$, and $M_{Pl}=NM$. From \eq~(\ref{eq:discretized}) it is evident
that we can now equally well drop on all the coordinates $x_i$ the
index $i$ by making the identification $x^\mu_i\equiv x^\mu$ and
describe on all sites $i$ the positions using only a single universal
coordinate system with coordinates $x^\mu$.

Following the effective
field theory approach advanced in
\Refs~\cite{Arkani-Hamed:2002sp,Arkani-Hamed:2003vb,Schwartz:2003vj},
we can parameterize the link fields $Y^\mu_{ji}$ in terms of small
deviations from $x^\mu$ as
\begin{equation}\label{eq:links}
 Y_{ji}^\mu(x_\mu)=x^\mu+\pi^\mu_{ji}(x_\mu)=x^\mu+
A^\mu_{ji}(x_\mu)+\partial^\mu\phi_{ji}(x_\mu),
\end{equation}
where $\pi^\mu_{ji}(x_\mu)$ behaves in the two-site model defined by
$GC_i$ and $GC_j$ as a Nambu--Goldstone boson, which is eaten to
provide the longitudinal component of a massive graviton
\cite{Arkani-Hamed:2002sp}. In \eq~(\ref{eq:links}), the
Nambu--Goldstone bosons $\pi^\mu_{ji}(x_\mu)$ have been rewritten in
terms of vector fields $A^\mu_{ji}$ and scalars $\phi_{ji}$, where the
$\phi_{ji}$ provide the longitudinal components of the massive
gravitons. In fact, transforming to the stationary (or unitary) gauge
$Y^\mu_{ji}(x^\mu)= x^\mu$, we observe that the ``hopping'' terms in
\eq~(\ref{eq:discretized}) produce graviton mass terms of the
Fierz--Pauli type \cite{Fierz:1939ix}. By analogy with the Eguchi--Kawai
model, the model with two discrete gravitational extra dimensions will
therefore lead to a multi-graviton theory, containing one zero-mode
graviton and a phonon-like spectrum of massive gravitons with masses
of order $\sim 1/(Na)$, which simulates in the IR a linear tower of KK
excitations.

Now, let us consider small fluctuations of the metrics
$g^i_{\mu\nu}=\eta_{\mu\nu}+h^i_{\mu\nu}$ around Minkowski space and
express the gravitons as $h^i=\sum_{n}e^{{\rm i}2\pi i\cdot
n/N}h_{n}$, where we have introduced the index $n\equiv(n_1,n_2)$ with
$n_1,n_2=1,2,\ldots$ and where we have suppressed for simplicity the
Lorentz indices. Similarly, we write the scalar modes of the
Nambu--Goldstone bosons as $\phi^{ij}=\sum_{n}e^{{\rm i}2\pi i\cdot
n/N}\phi_n^k$, where $k=1$ ($k=2$) if the vector $i-j$ points in the
$y_1$ ($y_2$) direction. Inserting \eq~(\ref{eq:links}) into
\eq~(\ref{eq:discretized}) and transforming to momentum space, we observe that
the action contains a part, which takes in the limit $n\ll N$ a form
of the type
\begin{eqnarray}\label{eq:momentumbasis}
 \mathcal{S}&=& \int d^4x\:N^2M^2\left[ h_n\left(\partial^2
+\frac{1}{a^2}\frac{n^2}{N}\right)h_{-n}\right.\nonumber\\
&+&\left.\frac{1}{a^2}\sum_{k=1,2}\left(
\frac{n_1+n_2}{N}h_n\partial^2\phi^k_{-n}+(\partial^2\phi^k_n)(\partial^2\phi^k_m)
(\partial^2\phi^k_{-n-m})\right)\right]+\dots,
\end{eqnarray}
where we have omitted terms involving the transverse spin-1 modes and
higher-order terms in the spin-2 excitations. After going to canonical
normalization, $h_n\rightarrow h_n'\equiv NMh_n$ and
$\phi^k_n\rightarrow{\phi^k_n}'\equiv\frac{(n_1+n_2)M}{a^2}\phi_n^k$,
we find from the $(\partial^2\phi)^3$ term in
\eq~(\ref{eq:momentumbasis}), when evaluated for the lowest lying
modes, the strong coupling scale
\begin{equation}\label{eq:strongcoupling}
\Lambda=(M_{Pl}N/R^4)^{1/5},
\end{equation}
which is similar to the corresponding value
in a single discrete extra dimension.  Now, in order to avoid
effects that are non-local in theory space
\cite{Arkani-Hamed:2003vb,Schwartz:2003vj}, the inverse lattice
spacing must always be smaller than the maximal short-distance cutoff
$\Lambda_{\rm max}\simeq\sqrt{M_{6D}/R}$. Assuming a
compactification scale of the order of $1/R\sim10^{-2}\:{\rm
eV}\sim(10\:\mu{\rm m})^{-1}$ and $M_{6D}\sim 10\:{\rm TeV}$, we thus
obtain a cutoff $\Lambda_{\rm max}\simeq 0.3\:{\rm MeV}$, which is by
many orders of magnitude smaller than $M_{6D}$. From a field theory
point of view, it would therefore be necessary to provide an
understanding of the hierarchy $a^{-1}\leq\Lambda_{\rm max}\ll
M_{6D}$, when the lattice spacing $a$ is dynamically generated in a
possible UV completion of the theory of massive gravitons.

This hierarchy will be even more enhanced in strong
gravitational fields, where the higher-order terms omitted in
\eq~(\ref{eq:momentumbasis}) can no longer be neglected. Generally,
for a macroscopic body with mass $M'$, the cutoff gets modified
as $\Lambda_{\rm max}\rightarrow \Lambda_{\rm max}'\simeq\Lambda_{\rm
max}(M_{Pl}/M')^{1/3}$ \cite{Arkani-Hamed:2002sp}. Currently,
the most precise Cavendish-like tests of Newton's law use test masses
of the order $M'\sim(1-10)\:{\rm g}$ (see, \eg,
\Ref~\cite{Hoyle:2004}). If we are only interested in local theory space
models of gravity that admit meaningful predictions in such laboratory
experiments, then the inverse lattice spacing must be even smaller than a
cutoff $\Lambda_{\rm max}\sim 1\:{\rm keV}$. This suggests to consider theory space formulations of
large extra dimensions which naturally generate small inverse
lattice spacings. Going to the extreme limit where the inverse lattice spacing becomes as small as a
few eV, such models would have the additional benefit that
the structure of theory space becomes potentially accessible in
low-energy experiments via neutrino oscillations, when a right-handed
neutrino is propagating in the latticized bulk.

Note that, to prevent the SM from getting strongly coupled at
$\sim\Lambda_{\rm max}$, we would require a
perturbative description of the gravitons above this scale. In absence of
such an UV completion for gravity,
however, we shall in the next section analyze instead a deconstructed
$U(1)$ gauge theory, which provides an UV completion for the theory of
massive gauge bosons. This has the advantage that one can readily formulate an
explicit mechanism which produces sub-mm lattice spacings from energy scales
above the lattice cutoff, while ensuring that the dynamics of the SM fermions
always remains perturbatively sensible.

\section{Deconstructed $U(1)$ on a disk}\label{sec:Model}
In this section, we will study the deconstruction of a $U(1)$ gauge theory in
a non-gravitational extra dimension  compactified on the boundary of a
two-dimensional disk. This theory space has been analyzed in the context of
supersymmetry breaking \cite{Arkani-Hamed:2001ed} and the
doublet-triplet splitting problem \cite{Witten:2001bf}. Various properties
of supersymmetric deconstructed $U(1)$ models have also been studied in
Ref.~\cite{Falkowski:2002vc}. On the boundary of the disk, sub-mm lattice
spacings can be generated from a hierarchy between the masses of the link
fields. The resulting coarse latticization is experienced by a 
right-handed neutrino, which propagates on the boundary of the disk and mixes with
the SM neutrinos located in the center. The deconstructed $U(1)$ acts on the SM fermions as a gauged $B-L$ symmetry that is broken at the TeV scale.

\subsection{Gauge sector}
\begin{figure}
\begin{center}
\includegraphics*[bb = 210 580 398 766,height=5cm]{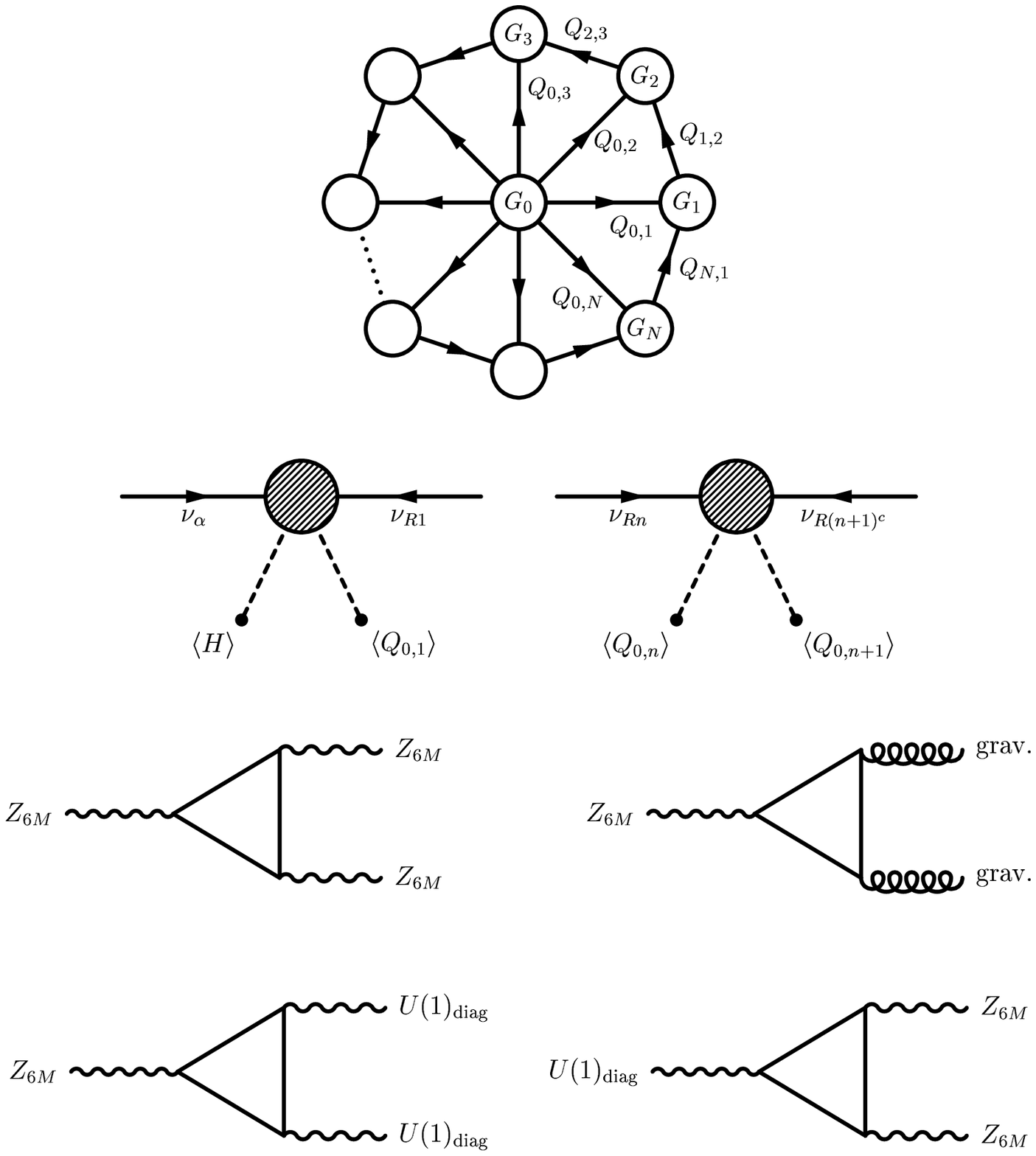}
\end{center}
\vspace*{-5mm}
\caption{\small{Moose diagram for the deconstructed $U(1)$ gauge
theory on a two-dimensional disk. Each circle corresponds to one
$U(1)_{i}\equiv G_{i}$ $(i=0,1,2,\ldots,N)$ gauge group. An arrow
pointing toward (outwards) a circle denotes a field with negative
(positive) charge under this group. The link fields $Q_{i,i+1}$ define
the boundary of the disk, while the radial links $Q_{0,i}$ connect the
gauge group in the center with the sites on the
boundary. }}\label{fig:disk}
\end{figure}

Consider the deconstruction of a $U(1)$ gauge theory, which is defined on the
boundary of a two-dimensional disk. Our
deconstructed theory is described in four dimensions by a
$U(1)^{N+1}\equiv\Pi_{i=0}^NU(1)_i$ product gauge group, where each
gauge group $U(1)_i$ corresponds to a site in theory space. The $N+1$
sites are connected by $2N$ scalar link variables $Q_{0,i}$ and
$Q_{i,i+1}$ $(i=1,2,\ldots,N)$, each of which carries under exactly two
neighboring gauge groups $U(1)_i$ and $U(1)_j$ the $U(1)_i\times
U(1)_j$ charges $(+1,-1)$ and transforms trivially under all the other
gauge groups. Specifically, for $i=1,2,\ldots ,N-1$, a link $Q_{i,i+1}$ is
charged as $(+1,-1)$ under the product group $U(1)_i\times U(1)_{i+1}$, while
the link $Q_{N,1}\equiv Q_{N,N+1}$ carries the $U(1)_N\times U(1)_1$
charges $(+1,-1)$. Thus, the sub-graph defined by the link fields
$Q_{i,i+1}$ has the geometry of a latticized circle. Each link
$Q_{0,i}$ ($i=1,2,\ldots,N$) carries the $U(1)_0\times U(1)_i$ charges
$(+1,-1)$, which leads to a ``non-local'' theory space, since any two
sites are connected through at most two links.  The theory space of
this model is conveniently represented by the ``moose'' \cite{geor86}
or ``quiver'' \cite{douglas:1996xx} diagram in \Fig~\ref{fig:disk}.
The gauge group $U(1)_0$ corresponds to the center of the disk, while the other
gauge groups $U(1)_{i\neq 0}$ define the sites on the boundary and are
connected by the ``boundary links'' $Q_{i,i+1}$. The ``radial links''
$Q_{0,i}$ connect the center to the sites on the boundary. We wish to
reiterate that, in contrast to the previous section, 4D gravity is simply added
to the 4D model, {\it i.e.}, our deconstructed extra-dimensional manifold is
non-gravitational.

It is useful to consider the global symmetry which corresponds to a
$2\pi/N$ rotation of the disk and acts on the link fields by
\begin{equation}\label{eq:ZN}
Z_N:\quad Q_{N,1}\rightarrow Q_{1,2},\quad Q_{0,N}\rightarrow Q_{0,1},\quad
Q_{i,i+1}\rightarrow Q_{i+1,i+2},\quad Q_{0,i}\rightarrow Q_{0,i+1},
\end{equation}
where $i=1,2,\ldots ,N-1$. Using the gauge degrees of freedom, we can
always establish an equivalence relation $Q_{1,2}\sim Q_{i,i+1}$, for
$i=1,2,\ldots,N$, which ``identifies'' the links on the
boundary. Now, for the lattice gauge field ``living'' on the disk, the
holonomy around each plaquette in \Fig~\ref{fig:disk} is trivial in
the lowest energy state. As a consequence, the Wilson lines will break
the symmetry $U(1)^{N+1}\times Z_N$ down to $U(1)_{\rm diag}\times F$,
where $U(1)_{\rm diag}$ is the diagonal subgroup of $U(1)^{N+1}$ and
$F$ can be taken as a diagonal product (linear combination) of a
$U(1)_0$ gauge transformation and the global $Z_N$ symmetry
\cite{Witten:2001bf}.

Initially, the scalar sector possesses a global $U(1)^{2N}$ symmetry,
which is broken by the gauge couplings, such that only a
$U(1)^{N+1}$ subgroup is preserved as the gauge symmetry of the
model. Wilson line breaking of the global symmetry leads to
 $2N$ Nambu--Goldstone boson
fields.  At the same time, the Wilson lines break also $N$ generators of the
$U(1)^{N+1}$ gauge symmetry, which produces $N$ massive spin-1 vector
states by eating $N$ of the Nambu--Goldstone boson fields via the Higgs
mechanism. Thus, we are left with $N$ classically massless
Nambu--Goldstone bosons in the low-energy theory, which can, however,
acquire a mass at the quantum level, since the large global $U(1)^{2N}$
symmetry is explicitly violated in the gauge sector.

The masses of the gauge bosons receive a contribution generated via the Higgs
mechanism from the kinetic terms of the link fields
$\sim\sum_{i=1}^N(|D_\mu Q_{0,i}|^2+|D_\mu Q_{i,i+1}|^2)$, where
$D_\mu Q_{j,i}=(\partial_\mu+{\rm i}g_jA_{j\mu}-{\rm i}g_iA_{i\mu})Q_{j,i}$ for
$(j,i)\in\{(0,i),(i,i+1)\}$ is the covariant derivative, in which 
$g_i$ and $A_{i\mu}$ denote the gauge coupling and the gauge boson of the group
$U(1)_i$. In the basis $(A_0^\mu,A_1^\mu,A_2^\mu,\ldots,A_N^\mu)$, the
$(N+1)\times (N+1)$ gauge boson mass squared matrix $M^2_A$, which results from these terms 
after spontaneous symmetry breaking (SSB), is therefore
\begin{equation}\label{eq:gaugemass}
 M^2_A=
g^2v^2\left(
\begin{matrix}
 N & -1 & -1 & -1 & \cdots\\
 -1 & 1 & 0 & 0 & \cdots\\
 -1 & 0 & 1 & 0 & \cdots\\
-1 & 0 & 0 & 1 & \cdots\\
\vdots & \vdots & \vdots & \vdots & \ddots
\end{matrix}
\right)+
g^2u^2
\left(
\begin{matrix}
 0 & 0 & 0 & 0 & \cdots\\
 0 & 2 & -1 & 0 & \cdots\\
 0 & -1 & 2 & -1 & \cdots\\
 0 & 0 & -1 & 2 & \cdots\\
\vdots & \vdots & \vdots & \vdots & \ddots
\end{matrix}
\right),
\end{equation}
where we have assumed, for simplicity, universal gauge couplings $g_i\equiv g$,
while $v$ and $u$ denote the universal VEV's $v\equiv\langle Q_{0,i}\rangle$
and $u\equiv\langle Q_{i,i+1}\rangle$, of the radial and the
boundary link fields, respectively. After diagonalization of the mass squared
matrix in \eq~(\ref{eq:gaugemass}), we arrive at the gauge boson mass spectrum
\begin{equation}\label{eq:gaugebosonmasses}
M_0^2=0,\quad
M_n^2=g^2v^2+4g^2u^2{\rm sin}^2\frac{\pi n}{N},\quad
M_N^2=(N+1)g^2v^2,
\end{equation}
where $n=1,2,\ldots,N-1$. We observe that this spectrum contains a zero mode,
which would correspond to an unbroken $U(1)_{\rm diag}$ (this symmetry will be
broken later, when we introduce fermions). In
\eq~(\ref{eq:gaugebosonmasses}), the tower of mass squares $M_n^2$,
where $n=1,2,\ldots,N-1$, reproduces for $n\ll N$ a spectrum of KK
modes of the order $(n/R)^2=(ngu/N)^2$, which has been shifted to
higher values by an additional universal contribution of the order
$(gv)^2$ provided by the radial links. One gauge boson with mass $M_N$
decouples at low energies for $N\rightarrow\infty$. Thus, in this
limit, the theory of massive gauge bosons becomes an effective
description of a latticized flat fifth dimension with an inverse
lattice spacing $u$ (generated by the boundary links) and one lattice
scalar (represented by the radial links), which acquires a VEV $v$ in
the 5D bulk. In the process, the link fields break the total $U(1)$
product gauge group down to the diagonal subgroup
$U(1)^{N+1}\rightarrow U(1)_{\rm{diag}}$, which can be further broken by
suitable scalar site variables in the center that acquire a VEV at the TeV
scale and will thus correspondingly modify the gauge boson spectrum in
Eq.~(\ref{eq:gaugebosonmasses}).

\subsection{Large lattice spacings}
In order to determine the actual vacuum structure in more detail, let
us now consider the scalar potential of the link fields in
isolation. The most general gauge invariant renormalizable scalar
potential of these fields then reads
\begin{eqnarray}\label{eq:potential}
V & = & \sum_{i=1}^{N}\Big{[}m^{2}|Q_{0,i}|^2+M^{2}|Q_{i,i+1}|^2
+ \mu Q_{0,i}Q_{i,i+1} Q_{0,i+1}^{\dagger} +
\mu^{\ast}Q_{0,i+1}Q_{i,i+1}^{\dagger} Q_{0,i}^{\dagger}\nonumber\\
&&+\frac{1}{2}\lambda_{1}|Q_{0,i}|^4 + \frac{1}{2}\lambda_{2}|Q_{i,i+1}|^4
+ \lambda_{3}^{ij}|Q_{i,i+1}|^2\sum_{j=1}^{N}|Q_{0,j}|^2
+ \lambda_{4}^{ij}|Q_{0,i}|^2\sum_{j\neq i}|Q_{0,j}|^2\nonumber\\
&&+\lambda_{5}^{ij}|Q_{i,i+1}|^2\sum_{j\neq i}|Q_{j,j+1}|^2
+(\lambda_{6}Q_{0,i}Q_{i,i+1}Q_{i+1,i+2}Q_{0,i+2}^{\dagger} + {\rm
h.c.})\Big{]},
\label{eq:seesawpotential}
\end{eqnarray}
where the parameters $m,M,$ and $\mu$ have mass dimension +1, while
$\lambda_{1},\lambda_{2},\lambda_3^{ij},\lambda_4^{ij},$ and
$\lambda_{5}^{ij}$ are dimensionless real parameters of order unity
and $\lambda_{6}$ is a complex-valued order unity coefficient. Note
that the potential $V$ is invariant under the global $Z_N$ symmetry in
\eq~(\ref{eq:ZN}). From the point of view of usual effective field
theories, the dimensionful parameters $m,M$, and $\mu$ may take any
value in the UV desert between $\sim1\:{\rm TeV}$ and $M_{Pl}$. We
will consider here the interesting case where these masses exhibit a
hierarchy $m\ll\mu\simeq M$, \ie, the boundary link fields are much
heavier than the radial link fields. To be specific, we assume that
the mass $m$ is close to the TeV scale, \ie, $m\simeq 1\:{\rm
TeV }$, while $\mu$ and $M$ are of the order the usual $B-L$ breaking scale
$\mu\simeq M \simeq M_{B-L}\simeq 10^{15}\:{\rm GeV}$. Note that an
understanding of the smallness of the parameter $m$ with respect to
$M_{B-L}$ may require mechanisms similar to those who give a solution to the
$\mu$-term problem in supersymmetric theories and will not be specifically
discussed here.

To explicitly minimize the scalar potential $V$, we shall now
make the simplifying assumption that the parameters $\mu$ and
$\lambda_6$ in \eq~(\ref{eq:seesawpotential}) are real. Actually, in a
supersymmetric case, holomorphy of the superpotential sets
$\lambda_6\rightarrow 0$ and one could rotate the phase of $\mu$ into
the Yukawa couplings of the neutrinos. Therefore, we argue that our
basic results concerning the magnitude of the VEV's will not be
significantly altered, when considering the more general case of
complex $\mu$ and $\lambda_6$. Now, taking $m^2<0$ and $\mu<0$, while
$M^2>0$, the potential $V$ in \eq~(\ref{eq:seesawpotential}) has an
extremum \cite{Bauer:2003mh}, which is given by $u\equiv\langle
Q_{i,i+1}\rangle$ and $v\equiv\langle Q_{0,i}\rangle$, for
$i=1,2,\ldots ,N$, where $u$ and $v$ are real and equal to
\begin{subequations}\label{eq:VEVs}
\begin{eqnarray}
u &\simeq&
\frac{m^{2}\mu}{2\left[\lambda_{1}+(N-1)\lambda_{4}\right]M^{2}-\mu^2},\label{eq:SmallVEVu}\\
v^{2} &\simeq& \frac{-m^{2}}{\lambda_{1} + (N-1)\lambda_{4}}
\left(1+\frac{u\mu}{m^2}\right),
\label{eq:LargeVEVv}
\end{eqnarray}
\end{subequations}
\ie, the boundary links and the radial links respectively acquire universal
VEV's. From \eqs~(\ref{eq:VEVs}) we observe that $u$ and $v^2$ become
$\sim 1/N$ suppressed in the large $N$ limit. However, let us now
consider the opposite situation, where $N\simeq\mathcal{O}(10)$, \ie,
the number of sites is kept moderate or small. In this case, we
observe that the choice of mass scales $m\simeq 1\:{\rm TeV}$
and $\mu\simeq M\simeq M_{B-L}\simeq 10^{15}\:{\textrm{GeV}}$ generates
for the boundary link fields $Q_{i,i+1}$ a small VEV of the order
$u\simeq 10^{-1}\:{\textrm{eV}}$, while the radial link fields
$Q_{0,i}$ acquire an unsuppressed TeV scale VEV $v\simeq 1\:{\rm TeV}$. In
other words, the model generates from mass
scales in the UV desert of conventional 4D theories an inverse lattice
spacing $u \sim (\mu{\rm m})^{-1}$ in the IR desert of large extra
dimensions. The suppression of $u$ due to the hierarchy $m \ll
\mu\simeq M$ is similar to the type-II seesaw mechanism \cite{typeII}
and, in fact, the structure of $V$ in \eq~(\ref{eq:seesawpotential})
can essentially be viewed as a replication of the model in
\Ref~\cite{Ma:1998dx}. It is the replication of gauge groups on the
boundary, which allows here to interpret $\sim u^{-1}$ as the sub-mm
lattice spacing of a deconstructed large extra dimension.

Note that our mechanism for the generation of sub-mm lattice spacings differs
from the model in Ref.~\cite{Bauer:2003mh} essentially in the choice of the dimensionful parameters $m$, $\mu$, and $M$. In Ref.~\cite{Bauer:2003mh}, they are of the orders $m\simeq\mu\simeq 10^2\:{\rm GeV}$ and $M\simeq 10^9\:{\rm GeV}$.
Having $\mu\simeq M$, however,
would be an automatic consequence in a supersymmetric version, where the tri-linear plaquette terms in Eq.~(\ref{eq:potential}) can emerge from the $F$-terms
of the superpotential. 

\subsection{Inclusion of fermions}
In our $U(1)^{N+1}$ model, we will first extend the three generations of SM
fermions by three fermions $N_1$, $N_2$, and $N_3$, which are singlets
under the SM gauge group $G_{SM}$. Then, the three fermion
generations are put on the center of the disk by assuming that they
carry nonzero $U(1)_0$ charges, but are singlets under the other gauge
groups $U(1)_{i\neq 0}$.  Note that the addition of matter fields as
site variables on the center leaves the $Z_N$ symmetry in
\eq~(\ref{eq:ZN}) unbroken.  We suppose that the leptons
$\ell_\alpha\equiv(\nu_\alpha,\:e_\alpha)^T$ and $e^c_\alpha$
($\alpha=1,2,3$ is the generation index) are charged under $U(1)_0$ as
$+1$ and $-1$, respectively, while the quark doublets $q_\alpha\equiv
(u_\alpha,\:d_\alpha)$ carry a $U(1)_0$ charge $-1/3$, and the
isosinglets $u^c_\alpha$, and $d^c_\alpha$ are given the $U(1)_0$
charges $+1/3$. The SM singlets $N_1$, $N_2$, and $N_3$, carry the
$U(1)_0$ charges $-4$, $-4$, and $+5$, respectively. Since the
$U(1)_0$ charges of the SM quarks and leptons are identical with their
$B-L$ quantum numbers, it is easily seen that the model will be free from
axial-vector \cite{axial-vector} and gauge-gravitational
\cite{gauge-gravitational} anomalies. Notice that this is slightly
different from the usual way of gauging $B-L$, where three
right-handed SM singlet neutrinos carry a $B-L$ charge $-1$
\cite{Marshak:1979fm}. With our charge assignment, however, the Yukawa
couplings of the active neutrinos to the fields $N_\alpha$ are
suppressed by many powers of $M_{Pl}$ and are thus negligible. Suitable SM
singlet scalar fields $S$ with nonzero $U(1)_0$ charges can then allow
renormalizable Yukawa couplings $\sim S N_\alpha N_\beta$ and break
the $U(1)_0$ symmetry around the TeV scale. (For a recent detailed analysis of
breaking $B-L$ at the TeV scale see, {\it e.g.}, Ref.~\cite{Anoka:2004vf}.)
The fields $N_\alpha$,
which were only introduced for the purpose of anomaly cancellation,
will then decouple below a TeV.

Next, we include a SM singlet fermion, which appears with respect to
the SM interactions as a right-handed neutrino propagating on the
boundary of the disk. In the deconstructed space, the bulk fermion is
represented by $N$ right-handed neutrinos $\Psi_i$ $(i=1,2,\ldots
,N)$, which are put as site variables on the boundary of the
disk. Here, $\Psi_i$ has a charge $-1$ under the group $U(1)_{i}$, but
is a singlet under the other gauge groups $U(1)_{j\neq i}$. In the
Weyl basis, we can decompose each field $\Psi_i$ as
$\Psi_{i}\equiv(\nu_{Ri},\:\overline{\nu^c_R}_i)^T$, where $\nu_{Ri}$
and $\overline{\nu^c_R}_{i}$ are two-component Weyl spinors.  The mass terms of the neutrinos will be discussed in the next section.

\section{Neutrino masses}\label{sec:neutrinomasses}
In this section, we will analyze the kinetic term of the latticized 
right-handed neutrino and its mixing with the SM neutrinos. Unwanted
higher-dimension operators can be eliminated by refining the triangulation
of the disk.

\subsection{Latticized right-handed neutrino}\label{sec:latticized}
The deconstructed model presented so far describes a non-local theory
space where any two sites are connected by at most two links. This is
in contrast to the continuum theory for neutrinos in large extra
dimensions \cite{Arkani-Hamed:1998vp,Dienes:1998sb}, where a local
interaction of a massless right-handed neutrino in the bulk leads to a
$\sim (M_{Pl})^{-1}$ suppressed coupling to the active
neutrinos. Moreover, the model is in its present form vector-like and
allows unprotected Dirac masses $\sim
M_{Pl}\nu_{Ri}\nu_{Ri}^c$. However, if we treat the latticized
right-handed neutrino as a massless Wilson fermion
\cite{Wilson:1974sk} propagating on the boundary of the disk, which has
lattice spacings in the sub-mm range, we can have only small Dirac
masses $\sim u\nu_{Ri}\nu_{Ri}^c$, where $u\sim(\mu{\rm m})^{-1}$ is the
inverse lattice spacing. In order to remedy this problem and make
contact with the 5D continuum theory in
\Refs~\cite{Arkani-Hamed:1998vp,Dienes:1998sb}, we introduce for each
site on the boundary of the disk a pair of scalars $\chi_n$ and
$\phi_n$ ($n=1,2,\ldots ,N$), which are $G_{SM}\times U(1)^{N+1}$
singlets, and assume a discrete $Z_{6M}$ symmetry ($M$ is an
appropriate integer) acting on the fields as
\begin{equation}\label{eq:Z6M}
Z_{6M}:\left\{
\begin{array}{ll}
\nu_{Rn}\rightarrow {\rm e}^{{\rm i} 2\pi(n+2)^2/M}\nu_{Rn},&
\nu_{Rn}^c\rightarrow {\rm e}^{-{\rm i}2\pi(n+1)^2/M}\nu^c_{Rn},\\
\chi_n\rightarrow{\rm e}^{-{\rm i}2\pi(2n+3)/M}\chi_n,&
\phi_n\rightarrow{\rm e}^{{\rm i}2\pi(2n+3)/(2M)}\phi_n,\\
Q_{0,n}\rightarrow{\rm e}^{{\rm i}16\pi/M}Q_{0,n},&
\ell_\alpha\rightarrow {\rm e}^{-{\rm i}2\pi/M}\ell_\alpha,\\
e^c_\alpha\rightarrow {\rm e}^{{\rm i}2\pi/M}e^c_\alpha,&
q_\alpha\rightarrow {\rm e}^{{\rm i}2\pi/(3M)} q_\alpha,\\
u^c_\alpha\rightarrow {\rm e}^{-{\rm i}2\pi/(3M)}u^c_\alpha,&
d^c_\alpha\rightarrow {\rm e}^{-{\rm i}2\pi/(3M)}d^c_\alpha,\\
N_{1,2}\rightarrow {\rm e}^{{\rm i}8\pi/M}N_{1,2},&
N_3\rightarrow {\rm e}^{-{\rm i}10\pi/M}N_3,
\end{array}
\right.
\end{equation}
where $n=1,2,\ldots ,N$ and $\alpha=1,2,3$ runs over all three
generations. Note that the left- and right-handed SM fermions carry
opposite charges under the $Z_{6M}$ symmetry, and hence, the Yukawa
couplings of the quarks and charged leptons will remain
unsuppressed. Furthermore, we note in \eq~(\ref{eq:Z6M}) that the
potential $V$ in \eq~(\ref{eq:seesawpotential}) remains invariant
under the $Z_{6M}$ symmetry.  In Appendix \ref{app:anomalies}, we show
that by adding extra fermions on the boundary the $Z_{6M}$ symmetry
can be promoted to a discrete gauge symmetry, which would be protected
from quantum gravity corrections \cite{Krauss:1988zc,Ibanez:hv}.  It
turns out that in the effective theory (ignoring the enlarged gauge
symmetry at high energies) all dangerous triangle diagrams would add
up to zero. It is interesting to note here, that the SM possesses an
anomaly-free $Z_6$ symmetry which can ensure nucleon stability for new
physics scales as low as $\sim 10^2\:{\rm GeV}$ \cite{Babu:2003qh}.

In the deconstructed theory, the action for neutrino masses which
includes all renormalizable interactions with $\mathcal{O}(1)$ Yukawa
couplings and the most general $U(1)^{N+1}$ invariant
dimension-five operators consistent with the discrete $Z_{6M}$ symmetry,
can now be written as
\begin{equation}\label{eq:Smass}
 \mathcal{S}_{\rm mass}=
\mathcal{S}_{\rm wilson}+\mathcal{S}_{\rm int}^{4D}+\mathcal{S}_{\rm dim 5},
\end{equation}
in which the different parts are given by
\begin{subequations}\label{eq:actions}
\begin{eqnarray}
 \mathcal{S}_{\rm wilson}&=&\int d^4x\sum_{n=1}^{N}u\nu_{Rn}
\Big(\frac{Q_{n,n+1}}{u}\nu^c_{R(n+1)}
-\frac{\chi_n}{u}\nu^c_{Rn}\Big)
+{\rm h.c.},\label{eq:Slink}\\
\mathcal{S}_{\rm int}^{4D}&=&
\int d^4x\frac{Y_\alpha}{M_f}\ell_\alpha\epsilon
HQ_{0,1}^\ast\nu_{R1}+{\rm h.c.},\label{eq:Smix}\\
\mathcal{S}_{\rm dim5}&=&
\int d^4x\sum_{n=1}^N\frac{Y_n}{M_f}
Q_{0,n}^\ast Q_{0,n+1}\nu_{Rn}\nu^c_{R(n+1)}
+{\rm h.c.}\label{eq:Sdim5},
\end{eqnarray}
\end{subequations}
where $\epsilon={\rm i}\sigma^2$ contracts the $SU(2)$ indices, while
$Y_\alpha$ and $Y_n$ are (complex) dimensionless $\mathcal{O}(1)$ Yukawa
couplings. The non-renormalizable operators $\mathcal{S}^{4D}_{\rm int}$ and $\mathcal{S}_{\rm dim5}$ are generated at the string or ``fundamental'' scale
$M_f\simeq(10^{17}-10^{18})\:{\rm GeV}$, where a value as low as
$M_f\simeq 10^{17}\:{\rm GeV}$ could be understood in M-theory \cite{Witten:1996mz}. In \eq~(\ref{eq:Slink}), let us assume that the $\chi_n$
acquire a universal VEV $\langle\chi_n\rangle=u$ $(n=1,2,\ldots ,N)$,
which is equal to the inverse lattice spacing $u$ defined by the
universal VEV's of the boundary links in \eq~(\ref{eq:SmallVEVu}). We
will comment on a possible origin of the order of this mass scale for
the VEV's of the $\chi_n$ later on. With the identification
$\langle\chi_n\rangle=u$, the action $\mathcal{S}_{\rm wilson}$ in
\eq~(\ref{eq:Slink}) takes the form of a Wilson-modified latticized 5D
kinetic term that describes the propagation of the right-handed neutrino
on the boundary of the disk, which is interpreted as a fifth
dimension. The fields $\nu_{R0}^c$ and
$\nu_{R(N+1)}^c$ in \eq~(\ref{eq:Slink})
are determined by $\nu_{R1}^c$ and $\nu_{RN}^c$ only
up to a discrete $Z_2$ ``gauge transformation'' reflecting the
topology of the disk. The $Z_2$ symmetry is associated with the
existence of non-trivial or twisted field configurations
\cite{Isham:1978xxx} for the latticized right-handed neutrino, which
are characterized by distinct spectra in the low-energy theory.  In
\eq~(\ref{eq:Slink}), we define $(\nu_{R0},\:\nu^c_{R0})=\pm
(\nu_{RN},\:\nu^c_{RN})$ and
$(\nu_{R(N+1)},\:\nu_{R(N+1)}^c)=\pm(\nu_{R1},\:\nu_{R1}^c)$, where
``$\pm$'' distinguishes between twisted ($-1$) and untwisted ($+1$)
fields. The effects of twisted field configurations in deconstruction
have been extensively discussed in \Ref~\cite{Hill:2002me}.

Upon using
the mechanism in \Sec~\ref{sec:Model} for generating small inverse
lattice spacings $u\sim (\mu{\rm m})^{-1}$, the latticized 5D kinetic term
$\mathcal{S}_{\rm wilson}$ leads then to an effective action for KK modes
\begin{equation}\label{eq:nuKK}
 \mathcal{S}_{\rm KK}=\int d^4x\sum_{n=1}^{N}u\nu_{Rn}
\Big(\nu^c_{R(n+1)}-\nu^c_{Rn}\Big)+{\rm h.c.}.
\end{equation}
In the basis spanned by $(\nu_{R1},\nu_{R2},\ldots,\nu_{RN})$ and
$(\nu^c_{R1},\nu^c_{R2},\ldots,\nu^c_{RN})$, the action
$\mathcal{S}_{\rm KK}$ in \eq~(\ref{eq:nuKK}) defines a Dirac mass
squared matrix $M^2$, which explicitly reads
\begin{equation*}
M^{2}=u^{2} \left(\begin{array}{rrrrr}
2 & -1 &  &  & -T\\
-1 & 2 & -1\\
 & \ddots & \ddots & \ddots\\
 &  & -1 & 2 & -1\\
-T &  &  & -1 & 2\end{array}\right),
\end{equation*}
where $T=\pm 1$ and the blank entries are all zero. The squared masses
$m_n^2$ of the fermions are found to be the eigenvalues of the matrix
$M^2$. Thus, we obtain for twisted $(T=-1)$ and untwisted $(T=+1)$
fields the mass spectra
\begin{equation}\label{eq:Wilsonspectra}
m_{n}^{2}=4u^2{\rm sin}^2\frac{(n-1/2)\pi}{N}
\quad ({\rm twisted}),\qquad 
m_{n}^{2}=4u^2{\rm sin}^2\frac{(n-1)\pi}{N}\quad
({\rm untwisted}),
\end{equation}
where $u\sim (\mu{\rm m})^{-1}$ and $n=1,2,\ldots,N$. We hence observe that
$\mathcal{S}_{\rm KK}$ reproduces in the IR for $\nu_{Rn}$ and
$\nu_{Rn}^c$ always a tower of KK excitations with $\sim(\mu{\rm
m})^{-1}$ masses, which becomes for large $N$ indistinguishable from
the lightest KK modes of a right-handed bulk neutrino in sub-mm sized
continuum extra dimensions. Note in Eq.~(\ref{eq:Wilsonspectra}), that
a zero mode is absent for twisted fields.

In the above discussion, we require that the $\chi_n$ acquire the
small VEV $\langle\chi_n\rangle=u\sim (\mu{\rm m})^{-1}$ to allow the
interpretation of $\mathcal{S}_{\rm wilson}$ as the Wilson action for
a massless right-handed neutrino. This energy scale has been generated
for the VEV's of the boundary links by the mechanism in
\Sec~\ref{sec:Model}. Since the potential for the $\chi_n$ and
$\phi_n$ is qualitatively similar to the potential $V$ of the link fields
in \eq~(\ref{eq:seesawpotential}), a variation of this
mechanism can also produce the right energy scale for
$\langle\chi_n\rangle$, when $\chi_n$ and $\phi_n$ take the r\^oles of
the boundary and radial links, respectively. For this purpose, we
suppose that the $\phi_n$ have masses $m_\phi$ around the TeV scale
$m_\phi\sim 10^2\:{\rm TeV}$, whereas the $\chi_n$ have masses
$M_\chi$ of the order the Planck scale $M_\chi\sim
M_{f}$. Additionally, we take in the renormalizable
$Z_{6M}$-invariant interactions $\sim\tilde{\mu}_n\phi_n\phi_n\chi_n$
the dimensionful couplings $\tilde{\mu}_n$ to be $\tilde{\mu}_n\sim
M_{f}$. By the same arguments as in \Sec~\ref{sec:Model}, we find
that the $\chi_n$ can acquire a VEV in the range
$\langle\chi_n\rangle\simeq m_\phi^2/M_{f}\sim (\mu{\rm m})^{-1}$,
which is of the order the inverse lattice spacing $u$ in
\eq~(\ref{eq:SmallVEVu}).

\subsection{Non-renormalizable terms}
\begin{figure}
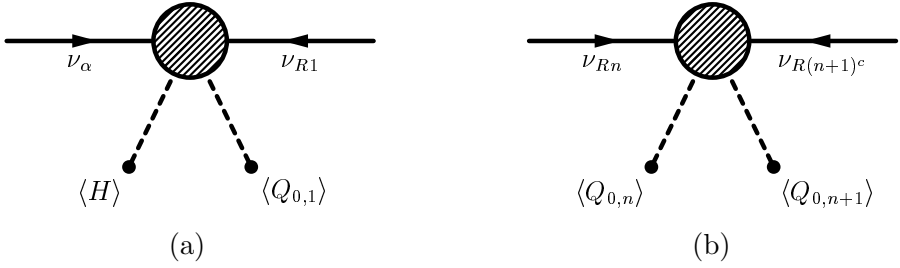

 \begin{center}
 \begin{tabular}{ccc}
 \includegraphics*[bb= 121 466 295 564,height=3cm]{figures.ps}&
 \hspace*{8mm}&
 \includegraphics*[bb= 315 466 489 564,height=3cm]{figures.ps}\\
 \small{(a)}&&\small{(b)}
\end{tabular} 
\vspace*{-2mm}
 \caption{\small{Non-renormalizable dimension-five operators generated
 at the fundamental scale $M_{f}$. When the link field $Q_{0,1}$ acquires a VEV $\langle Q_{0,1}\rangle\simeq 1\:{\rm TeV}$, this operator generates for $M_f\simeq 10^{17}\:{\rm GeV}$ a Dirac mass of the order $10^{-3}\:{\rm eV}$, which mixes the active neutrinos $\nu_\alpha$ with the tower of KK states via an
interaction with $\nu_{R1}$. In theory space, this dimension-five term
 corresponds to a Wilson line type effective operator, which connects
 the SM neutrinos $\nu_\alpha$ in the center with $\nu_{R1}$
 on the boundary of the disk via the link field
 $Q_{0,1}$. The operators in (b) generate Dirac masses of the order
$10^{-2}\:{\rm eV}$ for the right-handed neutrinos and are
neglected with respect to the nearest neighbor hopping terms of the order
$10^{-1}\:{\rm eV}$.}}\label{fig:Diracmass}
\end{center}
\end{figure}

The interaction of the active neutrinos $\nu_\alpha$ with the
right-handed neutrinos on the boundary of the disk is introduced by $\mathcal{S}_{\rm int}^{4D}$
in \eq~(\ref{eq:Smix}). In theory space, we identify the
dimension-five term \cite{Weinberg:1979sa,Wilczek:1979hc} in
$\mathcal{S}_{\rm int}^{4D}$ with a Wilson line type effective
operator, which connects the active neutrinos (in the center of the disk)
with $\nu_{R1}$ (on the boundary) via the link $Q_{0,1}$ (see
\Fig~\ref{fig:Diracmass}).
Let us now go to the basis where $\mathcal{S}_{\rm KK}$ in
\eq~(\ref{eq:nuKK}) is on diagonal form and consider the lowest lying
mass eigenstate $\nu_{R}'$ belonging to $\mathcal{S}_{\rm KK}$. After
setting $Q_{0,1}$ to its VEV $\langle Q_{0,1}\rangle=v\simeq
1\:{\rm TeV}$, the Wilson line type operator generates an
effective Yukawa interaction $\sim Y_\alpha\ell_\alpha\epsilon
H\nu'_{R}v/(\sqrt{N}M_f)$, which is suppressed by a factor $\sim
v/(\sqrt{N}M_f)$ with respect to the electroweak scale. For a string scale $M_f\simeq 10^{17}\:{\rm GeV}$ and small $N$ we thus obtain Dirac mass terms
$m_{D\alpha}\nu_a\nu'_R$, with Dirac masses
$m_{D\alpha}=Y_\alpha\langle H\rangle v/(\sqrt{N}M_f)\simeq 10^{-3}\:{\rm eV}$.

It is instructive to compare the effective Yukawa interaction generated by
$\mathcal{S}_{\rm int}^{4D}$ with the 5D ADD scenario. Here, a
right-handed bulk neutrino $\nu_R$ couples to the active neutrinos on
the SM brane through a local interaction
\cite{Arkani-Hamed:1998vp,Dienes:1998sb}
\begin{equation}\label{eq:braneinteraction}
 \mathcal{S}_{\rm int}^{5D}=\int d^4x
 \frac{Y^D_\alpha}{\sqrt{M_\ast}}\ell_\alpha(x)\epsilon H(x)
 \nu_R(x,y=0),
\end{equation}
where $y$ is the coordinate along the fifth dimension compactified on
a circle with circumference $2\pi R$, the coefficients $Y^D_\alpha$ are
dimensionless $\mathcal{O}(1)$ Yukawa couplings, and the SM lepton
doublets $\ell_\alpha$ as well as the Higgs doublet $H$ are 4D fields
trapped at $y=0$ on the SM brane. Note that while $\nu_{R1}$ in
$\mathcal{S}_{\rm int}^{4D}$ has mass dimension 3/2, the 5D fermion
$\nu_R$ in $\mathcal{S}_{\rm int}^{5D}$ has mass dimension 2. After
expanding $\nu_R$ in KK modes as $\nu_R(x,y=0)=(2\pi
R)^{-1/2}\sum_n\nu_{Rn}(x)$ and using the relation $2\pi
R=M_{Pl}^2/M_\ast^3$, it is seen that the interaction in
\eq~(\ref{eq:braneinteraction}) gives rise to a Dirac type coupling
$\sim Y^D_\alpha\ell_\alpha\epsilon H\nu_{R0}M_\ast/M_{Pl}$ between
the active neutrinos and the zero mode $\nu_{R0}$, which is
$M_\ast/M_{Pl}$ suppressed.  Since $v\sim M_\ast\sim 1\:{\rm TeV}$, we
thus find that, in the limit of coarse latticizations
$\sqrt{N}\sim\mathcal{O}(1-10)$, the couplings between the active and
the right-handed neutrinos generated by $\mathcal{S}_{\rm int}^{4D}$
and $\mathcal{S}_{\rm int}^{5D}$ become suppressed by factors of
similar orders $v/(\sqrt{N}M_f)\sim M_\ast/M_{Pl}$. However,
despite this numerical coincidence, the two models differ in an
interesting way: while the smallness of the Dirac type coupling in
$\mathcal{S}_{\rm int}^{5D}$ emerges from a volume suppression factor
(\ie, from the large number of KK modes below $M_\ast$), the small Dirac mass generated by $\mathcal{S}_{\rm int}^{4D}$
is rather a result of the separation between the site where the SM fermions are located and the boundary of the disk as compared to the length scale $M_f^{-1}$.

\begin{figure}
 \begin{center}
 \includegraphics*[bb= 187 531 426 766,height=6.2cm]{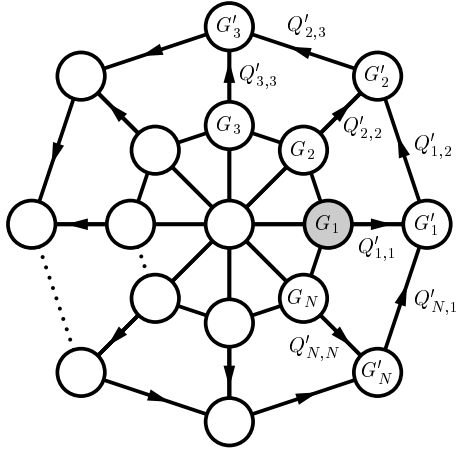}\\
 \vspace*{-2mm}
 \caption{\small{Spider web theory space for the suppression of unwanted
 higher-dimension operators. The inner disk is
  defined like in Fig.~\ref{fig:disk} and the latticized right-handed
 neutrino propagates on the outer circle formed by $N$ additional $U(1)$
gauge groups $U(1)_i'\equiv G_i'$ ($i=1,2,\ldots ,N$). The SM fermions are
placed on the site corresponding to $G_1$ (gray site). The link fields
$Q'_{i,i+1}$
on the outer circle have a mass of the order $10^{12}\:{\rm GeV}$,
whereas all other link fields have masses of the order $10^2\:{\rm TeV}$. Due
to the plaquette-structure near the boundary involving four links, the
operators analogous to Fig.~\ref{fig:Diracmass} (b) are suppressed by an
extra factor $\sim 10^{-12}$.}}\label{fig:spiderweb}
\end{center}
\end{figure}

The dimension-five operators contained in $\mathcal{S}_{\rm dim5}$ in
Eq.~(\ref{eq:Sdim5}) give for $M_f\simeq 10^{17}\:{\rm GeV}$ rise to Dirac
mass terms between $\nu_{Rn}$ and $\nu_{R(n+1)}^c$ that are of the order
$10^{-2}\:{\rm eV}$ (see Fig.~\ref{fig:Diracmass}). However, since
$\nu_{R1}$ is the only right-handed neutrino which couples ``directly''
(at the non-renormalizable level) to the active neutrinos, we may
treat the terms in $\mathcal{S}_{\rm dim5}$ as subleading corrections to
$\mathcal{S}_{\rm wilson}$, which gives Dirac masses of the order $10^{-1}\:{\rm eV}$, and ignore them in the following discussion. Let us now, instead,
consider a more attractive possibility to suppress the unwanted
non-renormalizable terms of the type shown in Fig.~\ref{fig:Diracmass}(b)
by making only use of the plaquette-structure of the model.
For this purpose, we will assume that the non-local theory space
introduced in Sec.~\ref{sec:Model} is actually part of a larger
``spider web theory space'' \cite{Arkani-Hamed:2001ed} as shown in
Fig.~\ref{fig:spiderweb}. 
This theory space is obtained from the disk in Fig.~\ref{fig:disk}
by adding $N$ extra $U(1)$ gauge groups
$U(1)_i'$ ($i=1,2,\ldots ,N$), which gives the
total gauge group $U(1)^{2N+1}\equiv
\Pi_{i=0}^NU(1)_i\times\Pi_{j=1}^NU(1)_j'$. Each pair of factors
$U(1)_i'$ and $U(1)_i$ is connected by a link field $Q_{i,i}'$ which is
charged under $U(1)_i\times U(1)_i'$ as $(+1,-1)$ and is a singlet under the other gauge groups.
Two neighboring groups $U(1)_i'$ and $U(1)_{i+1}'$, where $i\sim i+N$,
are connected by a boundary link field $Q_{i,i+1}'$, that is charged as
$(+1,-1)$ under $U(1)'_i\times U(1)'_{i+1}$ and transforms trivially under the
other gauge groups.

In analogy with Sec.~\ref{sec:Model}, we suppose that
the latticized right-handed neutrino propagates in Fig.~\ref{fig:spiderweb} on
the outer circle defined by the links $Q'_{i,i+1}$. Contrary to the previous
non-local theory space example, however, the SM fermions are now placed on the site associated with $G_1$ on the inner circle (see Fig.~\ref{fig:spiderweb}). The $U(1)_1$ charge assignment of the SM fermions is similar to that in Sec.~\ref{sec:Model} with $U(1)_0$ replaced by $U(1)_1$. Next, we suppose that the boundary
link fields $Q'_{i,i+1}$ on the outer circle have a common mass of the
order of an intermediate scale $10^{12}\:{\rm GeV}$ and positive mass squares.
The remaining link fields $Q_{0,i}$, $Q_{i,i+1}$, and $Q'_{i,i}$, on the
other hand, are supposed to have masses of the order $10^2\:{\rm TeV}$ and
negative mass squares. By the same arguments as in Sec.~\ref{sec:Model}, we
then find that the corresponding scalar potential is extremized for VEV's 
$\langle Q'_{i,i+1}\rangle\simeq 1\:{\rm eV}$, while all other link fields
have VEV's of the order $\langle Q_{0,i}\rangle\simeq\langle Q_{i,i+1}\rangle
\simeq\langle Q_{i,i}'\rangle\simeq 10^2\:{\rm TeV}$. The separations of the
sites on the outer circle are therefore in the sub-mm range, whereas the
inverse lattice spacings between all other neighboring sites are of the order
$10^{2}\:{\rm TeV}$. As a result, the active neutrinos couple to the
latticized neutrino on the outer circle via an analog of the operator
$S^{4D}_{\rm int}$ in Eq.~(\ref{eq:Smix}), where only $Q_{0,1}$ has been
replaced by $Q_{1,1}'$. More importantly, the ``dangerous'' dimension-five
operators of the type shown in Fig.~\ref{fig:Diracmass} (b) are now replaced
by dimension-six operators as
$\nu_{Rn}\nu_{R(n+1)}^cQ_{0,n}^\ast Q_{0,n+1}/M_f\rightarrow
\nu_{Rn}\nu_{R(n+1)}^c{Q_{n,n}'^\ast}Q_{n,n+1}Q_{n+1,n+1}'/M_f^2$.
As a consequence of the plaquette-structure near the boundary of the spider
web theory space, the unwanted higher-dimension operators are suppressed by an
extra factor $\sim10^{-12}$ and become therefore irrelevant. We thus see,
that the model in Sec.~\ref{sec:latticized} is completely
reproduced for a fundamental scale $M_f\simeq M_{Pl}\simeq 10^{18}\:{\rm GeV}$, but now without the
unwanted higher-dimension terms of the sort given in Eq.~(\ref{eq:Sdim5}).

Up to now, we have been considering the generation of Dirac neutrino
masses in a model, where $B-L$ is preserved by the link fields. In
order to understand recent solar \cite{solar}, atmospheric
\cite{atmospheric}, reactor \cite{KamLAND}, and accelerator \cite{K2K}
neutrino data we assume -- contrary to the usual type-I seesaw
mechanism \cite{typeI} -- that $B-L$ is broken at the TeV scale. One
possibility is here provided by versions of the Babu--Zee model
\cite{Babu-Zee}, which can be easily implemented in our model to
generate radiatively Majorana neutrino masses locally on the site where the SM fermions are located. However, in what follows, we will not further
specify the detailed mechanism which generates the Majorana masses of
the usual neutrinos. Instead, we will always assume the presence of
suitable Majorana mass terms and concentrate in the deconstructed
$U(1)$ model on the mixing of the SM neutrinos with the KK modes that
is introduced by the Dirac neutrino masses.

\section{Mixing with Kaluza--Klein modes}\label{sec:bulkmodes}

In this section, we will consider the neutrino mass and mixing terms
of our model by specializing to the simplifying case of only one
single active neutrino $\nu$ coupling to the latticized right-handed
neutrino, which we treat as a Wilson fermion. From the action in
\eq~(\ref{eq:Smass}), we thus obtain in this case after SSB for the
relevant neutrino mass and mixing terms the action density
\begin{equation}\label{eq:mass+mixing}
 \mathcal{L}^\nu_{\rm m}=m_\nu\nu\nu+\sqrt{N}m_D\nu\nu_{R1}
+u\nu_{RN}(T\nu^c_{R1}-\nu^c_{RN})  
+u\sum_{n=1}^{N-1}\nu_{Rn}(\nu^c_{R(n+1)}-\nu^c_{Rn})
+{\rm h.c.},
\end{equation}
where the small inverse lattice spacing $u\sim (\mu {\rm m})^{-1}$ has
been generated by the mechanism in \Sec~\ref{sec:Model} and the
parameter $T=\pm 1$ describes a twisted/untwisted right-handed
neutrino. In \eq~(\ref{eq:mass+mixing}), the Dirac mass type coupling
$\sqrt{N}m_D=M_f^{-1}\langle H\rangle\langle Q_{0,1}\rangle\simeq
10^{-2}\:{\rm eV}$ arises from the higher-dimension operator
shown in \Fig~\ref{fig:Diracmass} (a) (or the analogous term in the extension to
spider web theory space with $Q_{0,1}$ replaced by $Q_{1,1}'$,
see Sec.~\ref{sec:neutrinomasses}), while the Majorana mass
$m_\nu\simeq 10^{-2}\:{\rm eV}$ has some other origin and may, \eg,
emerge from a radiative mechanism as mentioned in
\Sec~\ref{sec:neutrinomasses}.  For
convenience, we have chosen for the second term in
\eq~(\ref{eq:mass+mixing}) a normalization factor $\sqrt{N}$, which is
related to the volume suppression factor in the corresponding 5D
continuum theory.

The action density $\mathcal{L}^\nu_{\rm m}$ in
\eq~(\ref{eq:mass+mixing}) translates into a $(2N+1)\times (2N+1)$
neutrino mass matrix from which we determine by diagonalization the
neutrino mass and mixing parameters for the different cases $N$
odd/even and $T$ twisted/untwisted. Since $m_\nu, \sqrt{N}m_D \ll u$,
it is useful to define the quantity $\epsilon\equiv \sqrt{N}m_D/u \ll
1$ as an expansion parameter in perturbation theory and diagonalize
the matrix $MM^{\dagger}$ in several steps. First, we bring the
latticized fermion kinetic term in \eq~(\ref{eq:mass+mixing}) on
(approximately) diagonal form by applying a transformation
$MM^{\dagger}\rightarrow U^TMM^\dagger U^\ast$ with a suitable unitary
matrix $U$. The mixing matrices $U$ for the different possible cases
are explicitly given in Appendix \ref{app:diagonalization}. For
definiteness, let us consider the case $T=-1$ and $N$ even, the other
cases follow then in similar ways. Transforming to momentum space with
respect to the latticized dimension defines a new basis
$(\nu,\hat\nu_{RN/2+1},\ldots,\hat\nu_{RN},\hat\nu^c_{R1},
\hat\nu^c_{R2},\ldots,\hat\nu^c_{RN})$, in which the resulting mass
squared matrix reads
\begin{equation}\label{eq:X}
{\footnotesize \hat{M}^2=u^2\left(\begin{array}{ccccccccccc} \lambda & \gamma & \ldots & \gamma & \gamma & a_{1}
 & \ldots & a_{N/2} & b_{N/2+1} & \ldots & b_{N} \\
\gamma & \lambda_{N/2+1}+\delta & \cdots & \delta & \delta  \\
\vdots & \vdots & \ddots & \vdots & \vdots \\
\gamma & \delta & \cdots &   \lambda_{N-1}+\delta & \delta\\
\gamma & \delta & \cdots & \delta & \lambda_{N}+\delta\\
a_{1} & & & & &   \lambda_{1}\\
\vdots & & & & & &  \ddots\\
a_{N/2} & & & & & & & \lambda_{N/2}\\
b_{N/2+1} & & & & & & & & \lambda_{N/2+1}  \\
\vdots & & & & & & & & & \ddots\\
b_{N} & & & & & & & & & & \lambda_{N} \\
\end{array}\right),}
\end{equation}
where the blank entries in this matrix are all zero. The nonzero
elements are given by
\begin{equation}\label{eq:gammadelta}
\gamma=\frac{\sqrt{2}m_{D}m_{\nu}}{u^2},\quad
\delta=\frac{2m_{D}^2}{u^2},\quad
a_{n}=\frac{\sqrt{2}m_{D}}{u}\sin{\frac{(2n-1)\pi}{N}},
\end{equation}
for $n=1,2,\ldots,N/2$ and
\begin{equation}\label{eq:b_{n}}
b_{n} = \frac{\sqrt{2}m_{D}}{u}
\left[-1+\cos{\frac{(2n-1)\pi}{N}}\right],
\end{equation}
for $n=N/2+1,N/2+2,\ldots,N$. In \eq~(\ref{eq:X}),
the masses $\lambda$ and and $\lambda_{n}$ are
\begin{equation}\label{eq:lambda}
\lambda=\frac{Nm_{D}^2+m_{\nu}^2}{u^2}\quad{\rm and}\quad
\lambda_{n}=4\sin^2{\frac{(n-1/2)\pi}{N},}
\end{equation}
for $n=1,2,\ldots,N $. The coefficients $\lambda_n$ show the
characteristic doubling of KK modes of a phonon-like spectrum, since
they satisfy $\lambda_{n}=\lambda_{N-n+1}$ for $n=1,2,\ldots,
N/2$. In \eq~(\ref{eq:X}), note that the KK states
$\hat\nu_{R1},\hat\nu_{R2},\ldots,\hat\nu_{RN/2}$ exhibit no Yukawa
interaction with the active neutrino $\nu$, and hence, decouple
completely from the SM interactions. Next, we apply to the basis
$(\nu,\hat\nu_{RN/2+1},\ldots,\hat\nu_{RN},\hat\nu^c_{R1},
\hat\nu^c_{R2},\ldots,\hat\nu^c_{RN})$ a sequence of rotations by
defining the orthogonal states
\begin{equation}\label{eq:rot2}
\tilde{\nu}_{Rn}^c\equiv s_{n}\hat\nu_{Rn}^c+c_{n}\hat\nu_{R(N-n+1)}^c\quad
{\rm and}\quad
{\nu_{Rn}^c}'\equiv c_{n}\hat\nu_{Rn}^c-s_{n}\hat\nu_{R(N-n+1)}^c,
\end{equation}
for $n=1,2,\ldots,N/2$. In \eq~(\ref{eq:rot2}), we have $s_{n} \equiv
\cos\frac{(n-1/2)\pi}{N}$ and $c_{n}
\equiv-\sin\frac{(n-1/2)\pi}{N}$. Crudely, this corresponds to
``rotating away'' in \eq~(\ref{eq:X}) half of the interactions $\sim
\nu\hat{\nu}^c_{Ri}$, which reduces the degeneracy of the problem from
three-fold to two-fold. In the new basis, the mass squared matrix is
given by
\begin{equation}\label{eq:Mtilde2}
\tilde{M}^2 = u^2\left(\begin{matrix}\lambda&\gamma&\ldots & \gamma & \gamma &d_{1} &
d_{2} &\ldots & d_{N/2}\\ \gamma & \lambda_{N/2}+\delta & \ldots & \delta & \delta \\ \vdots & \vdots &
\ddots & \vdots & \vdots \\ \gamma & \delta & \cdots & \lambda_{2}+\delta & \delta\\ \gamma & \delta & \cdots & \delta  & \lambda_{1}+\delta\\d_{1} & & & & & \lambda_{1}\\ d_{2} & & & & & & \lambda_{2}\\
\vdots & & & & & & & \ddots \\ d_{N/2} & & & & & & & & \lambda_{N/2}\\
\end{matrix}\right),
\end{equation}
where
\begin{equation}\label{eq:c_{n}}
d_{n} = \frac{2\sqrt{2}m_{D}}{u}\sin{\frac{(n-1/2)\pi}{N}}
\end{equation}
for $n=1,2,\ldots,N/2$. Here $\gamma$, $\delta$, $\lambda$, and
$\lambda_{n}$ are the same as in \eqs~(\ref{eq:gammadelta}) and
(\ref{eq:lambda}). In the continuum limit, we expect for large $N$ to
recover some relevant characteristics of a continuous large extra
dimension. In order to match onto the 5D continuum theory, we will
compare our model with the one given in
\Ref~\cite{Dvali:1999cn}. Since in this model Majorana masses are
absent, we assume in \eq~(\ref{eq:Mtilde2}) that $m_\nu\rightarrow 0$,
which implies that $\gamma\rightarrow 0$. Furthermore, the matrix
elements $\delta$ are small in comparison with the quantities
$\lambda_n$ and can therefore be neglected when calculating to lowest
order. This means that the $N/2$ states
$\tilde{\nu}_{R1}^c,\tilde{\nu}_{R2}^c,\ldots, \tilde{\nu}_{RN/2}^c$
spanning in \eq~(\ref{eq:Mtilde2}) the top-left $N/2\times N/2$
submatrix with entries $\lambda_i+\delta$ ($i=1,2,\ldots ,N/2$) on the
diagonal, decouple from $\nu$. Consequently, we end up with just one
KK tower of $N/2$ states ${\nu_{Rn}^c}'$ $(i=1,2,\ldots ,N/2)$, which
span the last $N/2$ rows and columns of $\tilde{M}^2$ in
\eq~(\ref{eq:Mtilde2}). The remaining entries in $\tilde{M}^2$ become
for $n\ll N$ asymptotically equal to
\begin{equation}\label{eq:d2_{n}}
u^2d_{n}\rightarrow\frac{\sqrt{2}m_{D}}{R}(n-1/2),\quad
u^2\lambda \rightarrow Nm_{D}^2,\quad
u^2\lambda_{n}\rightarrow \frac{1}{R^2}(n-1/2)^2,
\end{equation}
where we have used the fact that $u=N/(2\pi R)$. We will match our
model onto the 5D continuum theory by setting $m_D\equiv y\langle
H\rangle M_\ast/M_{Pl}$, where $y$ is some $\mathcal{O}(1)$ Yukawa
coupling in the ADD scenario.  With this identification, our model
reproduces for the case $N$ odd and $T$ untwisted (see
\Tab~\ref{tab:cases}) in the IR exactly the effective neutrino mass
squared matrix of the 5D continuum theory for neutrino oscillations in
extra dimensions as discussed in \Ref~\cite{Dvali:1999cn}.
\begin{table}
\begin{center}
\begin{tabular}{|c|c|c|c|c|c|}
\hline
$T$&$N$&$n_{\rm max}$& $d_{n=1,2,\ldots,n_{\rm max}-1}$ & 
$d_{n_{\rm max}}$ & $\lambda_{n=1,2,\ldots,n_{\rm max}}$\\
\hline\hline
$+1$&even&$N/2+1$&
$\frac{2\sqrt{2}m_D}{u}\:{\rm sin}\frac{(n-1)\pi}{N}$
&$-\frac{2m_D}{u}$&$4\:{\rm sin}^2\frac{(n-1)\pi}{N}$\\
$+1$&odd&$N/2+1/2$&$\frac{2\sqrt{2}m_D}{u}\:{\rm sin}\frac{(n-1)\pi}{N}$&
$\frac{2\sqrt{2}m_D}{u}\:{\rm sin}\frac{(N-1)\pi}{2N}$
&$4\:{\rm sin}^2\frac{(n-1)\pi}{N}$\\
$-1$&even&$N/2$&$\frac{2\sqrt{2}m_D}{u}\:{\rm sin}
\frac{(n-1/2)\pi}{N}$&
$\frac{2\sqrt{2}m_D}{u}\:{\rm sin}\frac{(N-1)\pi}{2N}$
&$4\:{\rm sin}^2\frac{(n-1/2)\pi}{N}$\\
$-1$&odd&$N/2+1/2$&$\frac{2\sqrt{2}m_D}{u}\:{\rm sin}\frac{(n-1/2)\pi}{N}$&$-\frac{2m_D}{u}$&$4\:{\rm sin}^2\frac{(n-1/2)\pi}{N}$\\
\hline
\end{tabular}
\caption{\small Relevant entries in the neutrino mass matrices of the
type shown in \eq~(\ref{eq:Mtilde2}) in the limit $\gamma\rightarrow
0$ for the different cases $T$ twisted/untwisted and $N$ odd/even. The first and
second line reproduce for $n\ll N$ exactly the 5D continuum theory
results of \Ref~\cite{Dvali:1999cn} and the third line corresponds to
the example considered in the text. In all cases, the parameter
$\lambda$ is as given in \eq~(\ref{eq:lambda}).}\label{tab:cases}
\end{center}
\end{table}

Next, we will diagonalize $\tilde{M}^2$ by using two-fold degenerate
Rayleigh--Schr\"odinger perturbation theory. 
We start by rewriting the mass squared matrix $\tilde{M}^2$ as
$\tilde{M}^2 = M_{0}^2+\epsilon M_{1}^2$, where $M_{0}^2$ is a
diagonal matrix, $M_{1}^2$ is the perturbation matrix, and
$\epsilon=\sqrt{N}m_D/u\ll 1$ is the small expansion parameter. In
order for perturbation theory to be valid, we require that
$|\langle\phi_{ir}^{(0)}|\epsilon
M_{1}^2|\phi_{js}^{(0)}\rangle/(E_{i}^{(0)}-E_{j}^{(0)})|\ll1$ for
$i\neq j$, where $|\phi_{ir}^{(0)}\rangle $ denotes the zeroth order
eigenvector, $E_{i}^{(0)}$ the corresponding eigenvalue, and $r$ and $s$ are
the degeneracy indices. This means that we will require that
\begin{equation}\label{eq:constraint1}
\left|\frac{2\sqrt{2}m_{D}u\sin\frac{(n-1/2)\pi}{N}}{Nm_{D}^2+m_{\nu}^2-4u^2\sin^2\frac{(n-1/2)\pi}{N}}\right|\ll1
\quad{\rm and}\quad
\left|\frac{\sqrt{2}m_{D}m_{\nu}}{Nm_{D}^2+m_{\nu}^2-4u^2\sin^2\frac{(n-1/2)\pi}{N}}\right|\ll1.
\end{equation}
{}From these relations we note that perturbation theory will not be
valid for arbitrarily large $N$, since the denominator becomes
singular at some point when $Nm_{D}^2+m_{\nu}^2 \sim
4u^2\sin^2(n-1/2)\pi/N$. For the other cases, $T=-1$, $N$ odd and
$T=1$, $N$ odd/even, one obtains essentially the same
constraints. Now, we apply perturbation theory to this problem and
obtain the matrix that diagonalizes $\tilde{M}^2$ as a result. We
denote this matrix by $W^{(k)}$, where $k$ denotes the $k$th order in
perturbation theory.  Thus, the mixing matrix $V$ which relates the
original basis to the mass eigenstate basis via $MM^\dagger\rightarrow
V^TMM^\dagger V^\ast$ is given by
\begin{equation}
V=UP_{1}P_{2}\cdots P_{N/2}W^{(k)},
\end{equation} 
where $P_{n}$ are the rotation matrices associated with the state
redefinitions in \eq~(\ref{eq:rot2}), and $W^{(k)}$, as stated above,
is the matrix of eigenvectors of $\tilde{M}$ calculated to some order
$k$ in perturbation theory. The first row is what is of interest to
us, since it gives the relevant mixing angles of $\nu$ with the bulk
modes. It will be entirely determined by $W^{(k)}$. Thus, to lowest
order in perturbation theory, we find
\begin{equation}\label{eq:V2}
V = \left( \begin{matrix} 1 & \epsilon A_{1} & \cdots
  & \epsilon A_{N/2} & \epsilon B_{1} & \cdots & \epsilon B_{N/2}\\
* & & & \cdots & & & *\\
\vdots & & & \ddots & & & \vdots\\
* & & & \cdots & & & *\\
\end{matrix}\right),
\end{equation}
which is an $(N+1) \times (N+1)$ orthogonal matrix, where
\begin{equation}\label{eq:A_n}
A_{n}=\frac{m_{\nu}}{u\sqrt{8N}\left[\sin^2\frac{(n-1/2)\pi}{N}-\frac{\lambda}{4}\right]}
\quad {\rm and}\quad
B_{n}=\frac{\sin\frac{(n-1/2)\pi}{N}}{\sqrt{2N}\left[
\sin^2\frac{(n-1/2)\pi}{N}-\frac{\lambda}{4}\right]},
\end{equation}
for $n=1,2,\ldots, N/2$ and the elements denoted by $*$ are not
relevant in the following discussion. Note that one can diagonalize
$MM^{\dagger}$ in other ways then the one described above. For
example, one could have applied four-fold degenerate perturbation
theory directly to the matrix $MM^{\dagger}$. One can show that this
gives the same result for the final neutrino oscillation
probabilities. However, reducing the degeneracy makes the problem much
easier to handle.
\section{Neutrino oscillations}
\label{sec:neutrinooscillations}
Global analyses have well established that the standard active
three-flavor neutrino oscillations with mass squared differences of
the orders of magnitude $\Delta m_{21}^2 \simeq 8.1 \cdot
10^{-5}\:{\rm eV}^2$ and $\Delta m_{31}^2 \simeq 2.2 \cdot
10^{-3}\:{\rm eV}^2$ are in excellent agreement with neutrino
oscillation data (see, \eg, \Ref~\cite{Maltoni:2004ei}). However, the
KK modes of the latticized right-handed neutrino could provide a
sizable subdominant component in solar and atmospheric neutrino
oscillations, and thus, lead to new anomalies, which are in reach of
more precise ongoing or future neutrino oscillation experiments. In
this section, we will derive the corresponding neutrino oscillation
formulas for our deconstructed model. These will only be valid in the
regime where the mixing parameters satisfy the constraints in
\eq~(\ref{eq:constraint1}).  For phenomenologically allowed values of
the physical parameters, this means that the formulas below will in
general be valid for a low or moderate number of sites, $N\lesssim
10-100$.

\begin{figure}
\begin{center}
\includegraphics*[width=17 cm]{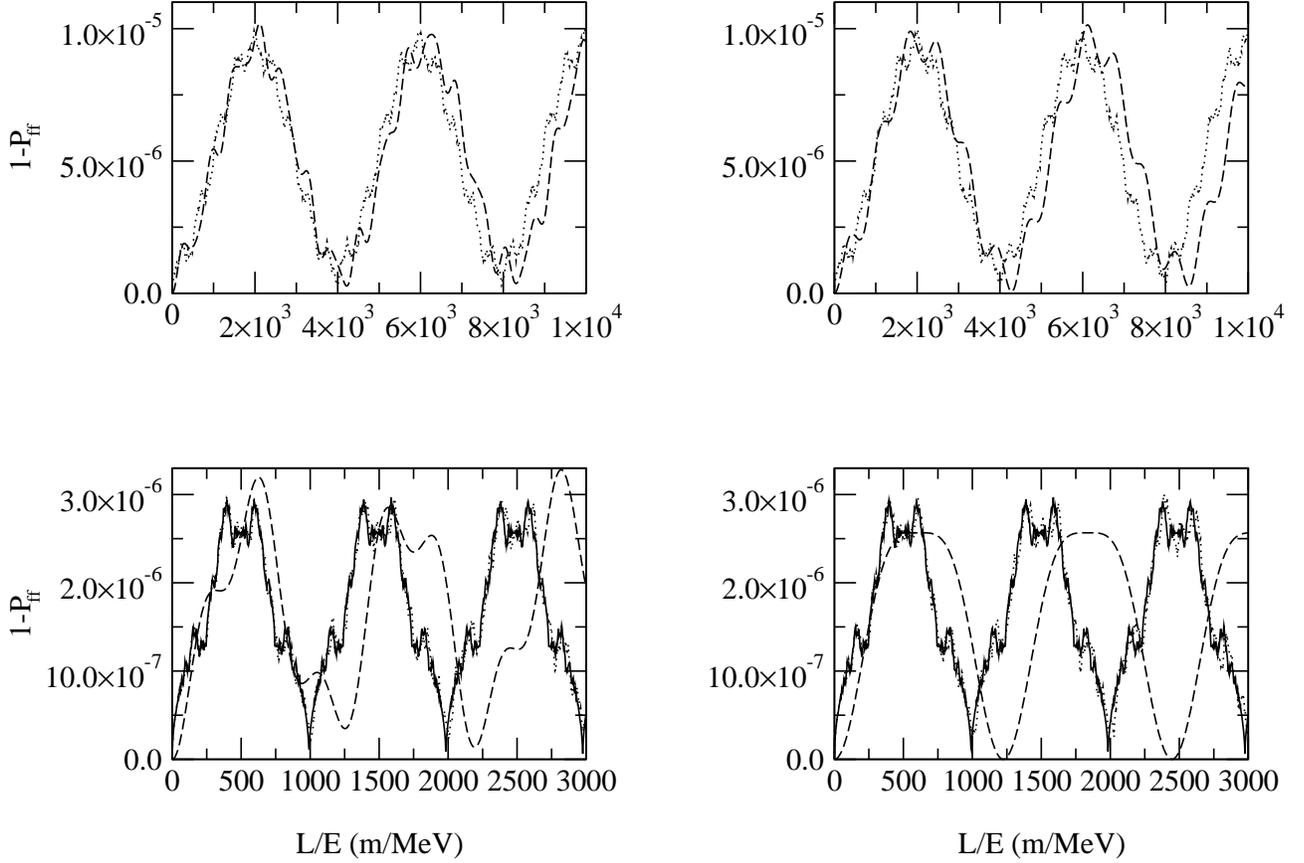}
\end{center}
\caption{\small{The neutrino transition survival probability $P_{ff}$ as a
function of $L/E$ for some choices of $N$ even or odd and $T= \pm
1$. Here we have set $R^{-1}=0.1\:{\rm eV}$, $m_{D}=5\cdot
10^{-5}\:{\rm eV}$, and $m_{\nu}=0\:{\rm eV}$. {\it Upper left panel:} $T=-1$ and $N$
odd for $N=5$ (dashed curve) and $N=55$ (dotted curve). {\it Upper right
panel:} $T=-1$ and $N$ even for $N=4$ (dashed curve) and $N=44$ (dotted
curve). {\it Lower left panel:} $T=1$ and $N$ odd for $N=5$ (dashed curve),
$N=55$ (dotted curve), and the continuum model (solid curve). {\it Lower
right panel:} $T=1$ and $N$ even for $N=4$ (dashed curve), $N=44$
(dotted curve), and the continuum model (solid
curve).}}\label{fig:plot1}
\end{figure}

\begin{figure}
\begin{center}
\includegraphics*[width=17 cm]{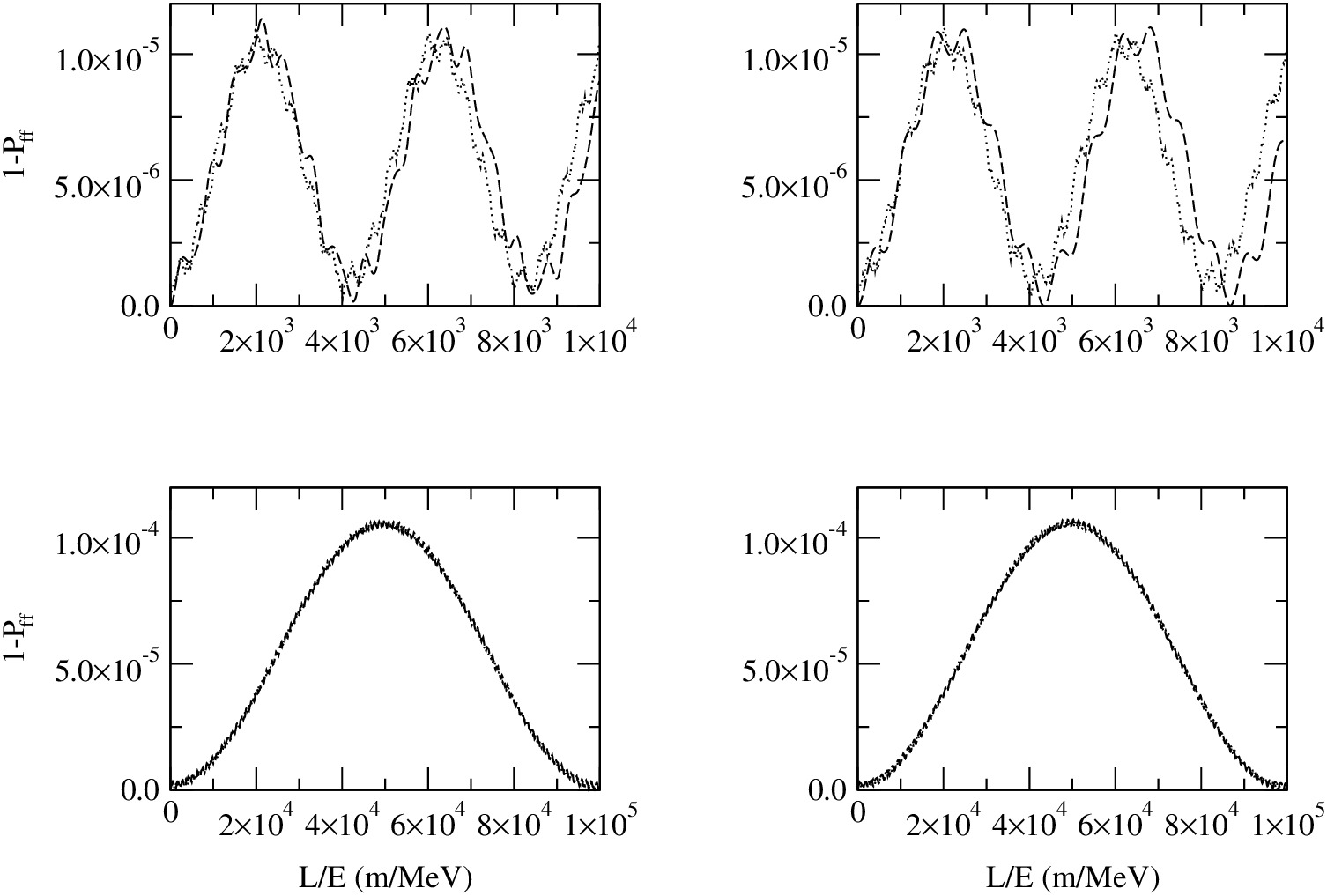}
\end{center}
\caption{\small{The neutrino transition survival probability $P_{ff}$ as a
function of $L/E$ for some choices of $N$ even or odd and $T= \pm 1$. Here we have set
$R^{-1}=0.1\:{\rm eV}$, $m_{D}=5\cdot 10^{-5}\:{\rm eV}$, and
$m_{\nu}=0.01\:{\rm eV}$. {\it Upper left panel:} $T=-1$ and $N$ odd for
$N=5$ (dashed curve) and $N=55$ (dotted curve). {\it Upper right panel:}
$T=-1$ and $N$ even for $N=4$ (dashed curve) and $N=44$ (dotted
curve). {\it Lower left panel:} $T=1$ and $N$ odd for $N=5$ (dashed curve)
and $N=55$ (dotted curve). {\it Lower right panel:} $T=1$ and $N$ even for
$N=4$ (dashed curve) and $N=44$ (dotted curve).}}
\label{fig:plot2}
\end{figure}

\begin{figure}
\begin{center}
\includegraphics*[width=17 cm]{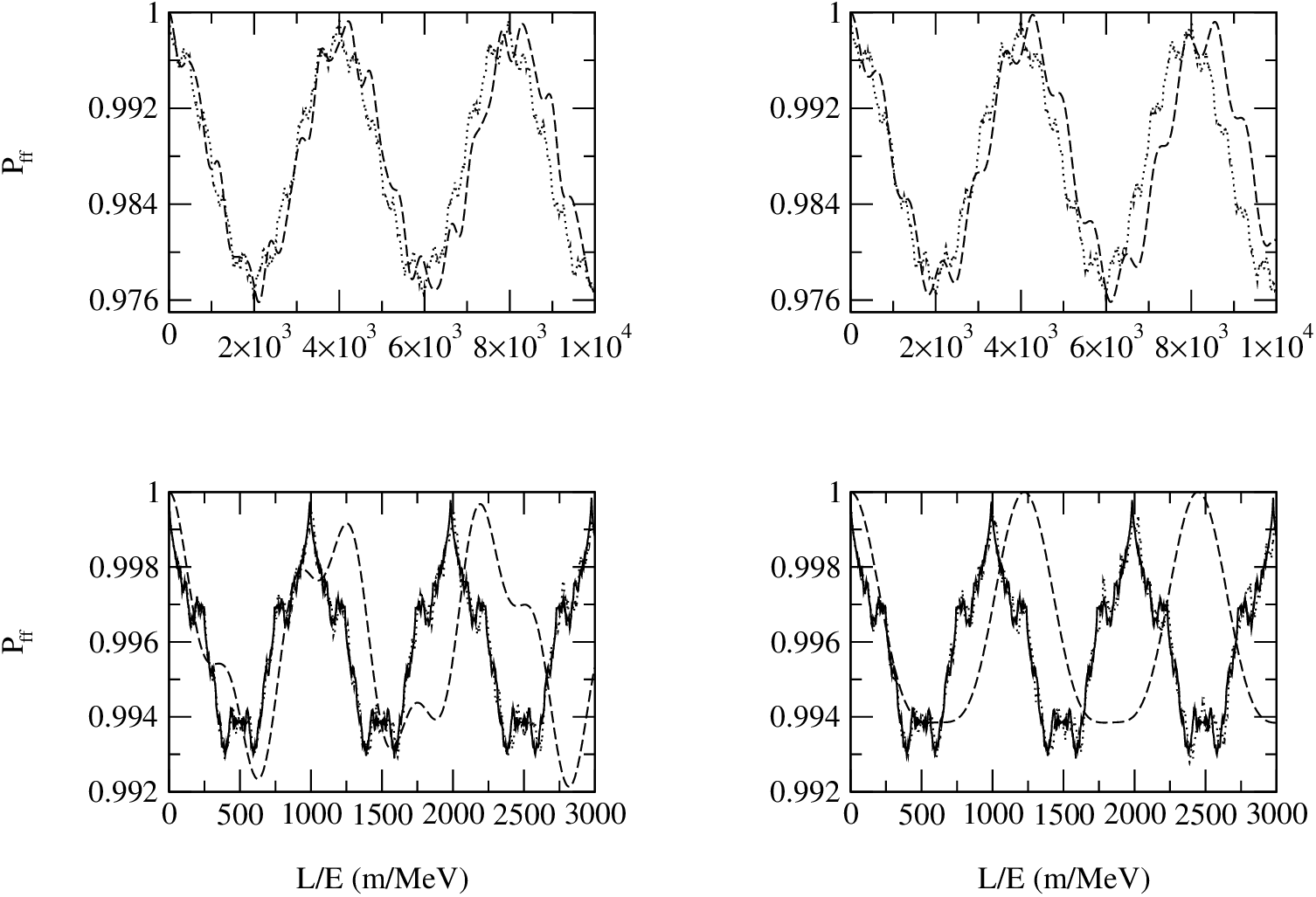}
\end{center}
\caption{\small{The neutrino transition survival probability $P_{ff}$ as a
function of $L/E$ for some choices of $N$ even or odd and $T= \pm 1$. Here we have set
$R^{-1}=0.1\:{\rm eV}$, $m_{D}=2.5\cdot 10^{-3}\:{\rm eV}$, and
$m_{\nu}=0\:{\rm eV}$. {\it Upper left panel:} $T=-1$ and $N$ odd for $N=5$ (dashed
curve) and $N=55$ (dotted curve). {\it Upper right panel:} $T=-1$ and $N$
even for $N=4$ (dashed curve) and $N=44$ (dotted curve). {\it Lower left
panel:} $T=1$ and $N$ odd for $N=5$ (dashed curve), $N=55$ (dotted
curve), and the continuum model (solid curve). {\it Lower right panel:} $T=1$ and $N$ even for $N=4$ (dashed
curve), $N=44$ (dotted curve), and the continuum model (solid curve).}}
\label{fig:plot3}
\end{figure}

\begin{figure}
\begin{center}
\includegraphics*[width=17 cm]{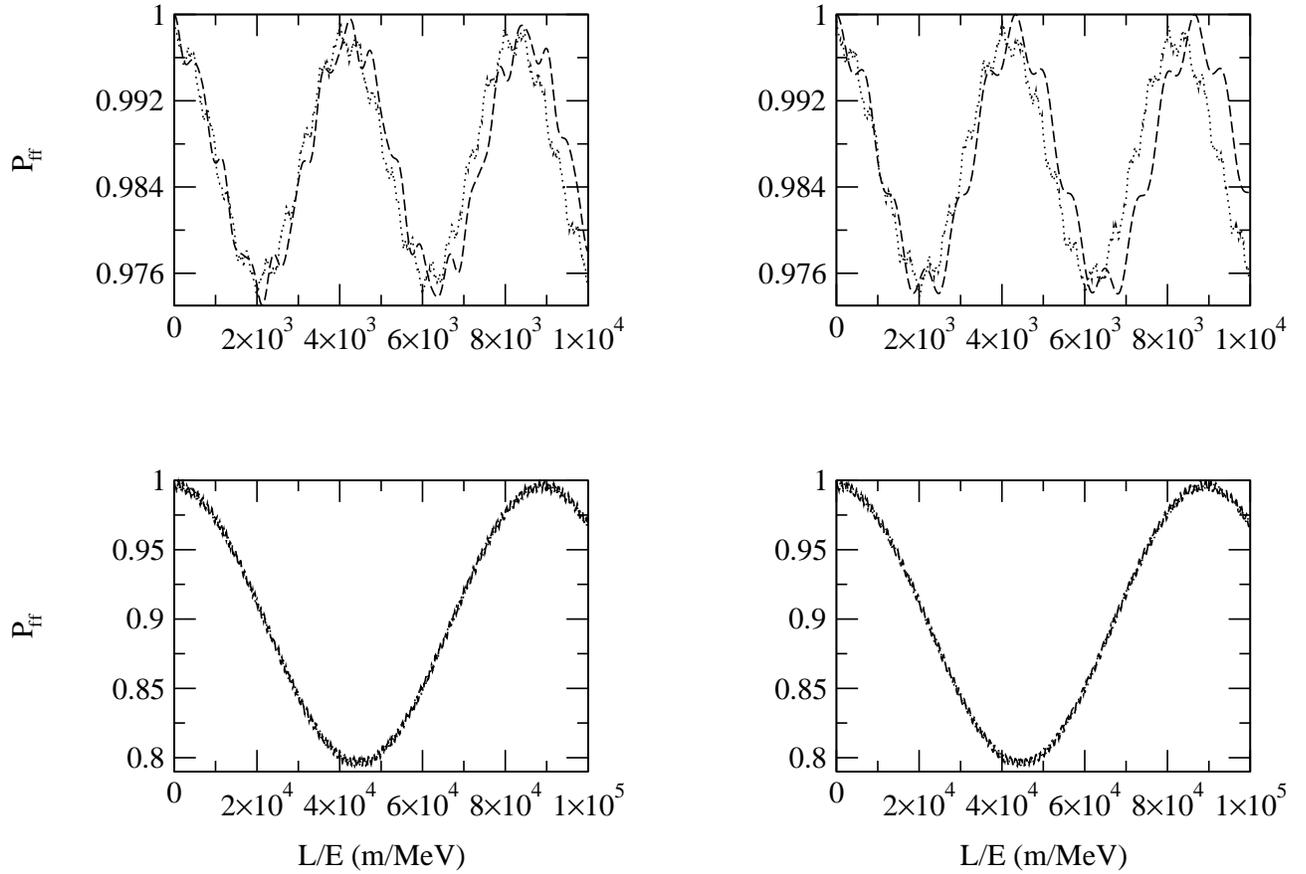}
\end{center}
\caption{\small{The neutrino transition survival probability $P_{ff}$ as a
function of $L/E$ for some choices of $N$ even or odd and $T =\pm 1$. Here we have set
$R^{-1}=0.1\:{\rm eV}$, $m_{D}=2.5\cdot 10^{-3}\:{\rm eV}$, and
$m_{\nu}=0.01\:{\rm eV}$. {\it Upper left panel:} $T=-1$ and $N$ odd for
$N=5$ (dashed curve) and $N=55$ (dotted curve). {\it Upper right panel:}
$T=-1$ and $N$ even for $N=4$ (dashed curve) and $N=44$ (dotted
curve). {\it Lower left panel:} $T=1$ and $N$ odd for $N=5$ (dashed curve)
and $N=55$ (dotted curve). {\it Lower right panel:} $T=1$ and $N$ even for
$N=4$ (dashed curve) and $N=44$ (dotted curve).}}
\label{fig:plot4}
\end{figure}

In order to describe the neutrino oscillations in our model, we can
write the flavor eigenstates as a linear combination of the mass
eigenstates using the mixing matrix in \eq~(\ref{eq:V2}). We find
that
\begin{equation}\label{eq:linearcomb}
|\nu_{f}\rangle = \frac{1}{K}\left(|\nu \rangle + \epsilon \sum_{n=1}^{N/2}
 A_{n}|\hat\nu_{n}\rangle + \epsilon \sum_{n=1}^{N/2} B_{n}|\nu_{n}\rangle \right).
\end{equation}
Here $|\nu_{f}\rangle$ denote a flavor eigenstate for some flavor
$f\in\{e,\mu,\tau\}$ and $|\nu\rangle$, $|\hat\nu_{n}\rangle$, and
$|\nu_{n}\rangle$ denote the mass eigenstates. We have also introduced
a normalization constant $K$, which follows from the condition
$|\langle\nu_{f}|\nu_{f}\rangle|^2=1$.  Thus, we find from
\eq~(\ref{eq:linearcomb}) that
\begin{equation}\label{eq:K}
K^2=1+\epsilon^2 \sum_{n=1}^{N/2} \left(A_{n}^2 + B_{n}^2 \right).
\end{equation}
Next, in the transition survival probability $P_{ff}\equiv
P(\nu_{f}\to\nu_{f})\equiv|\langle\nu_{f}|\nu_{f}(t)\rangle|^2$, the
time-evolved state $|\nu_{f}(t)\rangle$ is given by
\begin{equation}\label{eq:Ht}
|\nu_{f}(t)\rangle = \frac{1}{K}{\rm e}^{-{\rm i }\frac{\left (Nm_{D}^2+m_{\nu}^2\right) t}{2E}}\left(|\nu\rangle+\epsilon\sum_{n=1}^{N/2}A_{n}{\rm e}^{{\rm i}\phi_{n}}|\hat\nu_{n}\rangle+\epsilon \sum_{n=1}^{N/2}B_{n}{\rm e}^{{\rm i}\phi_{n}}|\nu_{n}\rangle \right),
\end{equation}
where the phases $\phi_{n}$ and the mass-squared eigenvalues $m_n^2$ equal
[\cf~\eq~(\ref{eq:Wilsonspectra})]
\begin{equation}\label{eq:phases}
\phi_{n}=\frac{(Nm_{D}^2+m_{\nu}^2-m_{n}^2)t}{2E}\quad{\rm and}
\quad m_{n}^2=4u^2\sin^2\frac{(n-1/2)\pi}{N},
\end{equation}
in which $E$ is the neutrino energy. Using \eqs~(\ref{eq:linearcomb})
and (\ref{eq:Ht}) gives for the case $T=-1$ and $N$ even
\begin{equation}\label{eq:P2}
P_{ff}= \frac{1}{K^4}\left|1+\epsilon^2 \sum_{n=1}^{N/2}(A_{n}^2+B_{n}^2){\rm e}^{{\rm i}\phi_{n}} \right|^2.
\end{equation}
For the other cases one finds the transition probability expressions
in similar ways, first starting by applying the matrix $U_{1}$ for the
case one considers and then by using a set of rotations similar to the
ones given by \eq~(\ref{eq:rot2}).  Next, one applies perturbation
theory and finds the mixing matrix from which the transition survival
probability expressions follows. Thus, for the case $T=-1$ and $N$ odd
we have
\begin{equation}\label{eq:PT-1O}
P_{ff}=\frac{1}{K^4}\left|1+\epsilon^2\left(\sum_{n=1}^{\frac{N-1}{2}}(A_{n}^2+B_{n}^2){\rm e}^{{\rm i}\phi_{n}}+\beta {\rm e}^{{\rm i}\phi_{\frac{N+1}{2}}} \right)\right|^2,
\end{equation}
where $$\beta=\frac{m_\nu^2}{Nu^2\left(4-\frac{Nm_{D}^2+m_{\nu}^2}{u^2}\right)^2}+\frac{4}{N\left(4-\frac{Nm_{D}^2+m_{\nu}^2}{u^2}\right)^2}.$$ 
Similarly, for $T=1$ and $N$ even we find
\begin{equation}\label{eq:PT1E}
P_{ff}=\frac{1}{K^4}\left|1+\epsilon^2\left(\alpha {\rm e}^{{\rm i}\phi_{1}}+\sum_{n=2}^{N/2}(A_{n}^2+B_{n}^2){\rm e}^{{\rm i}\phi_{n}}+\beta {\rm e}^{{\rm i}\phi_{\frac{N+2}{2}}}\right)\right|^2,
\end{equation}
where $$\alpha=\frac{m_{\nu}^2u^2}{N\left(Nm_{D}^2+m_{\nu}^2\right)^2}$$ and $\beta$ is the same as in \eq~(\ref{eq:PT-1O}).
Note the presence of a zero mode in the phase $\phi_{1}$.
Finally, for $T=1$ and $N$ odd we find
\begin{equation}\label{eq:PT1O}
P_{ff}=\frac{1}{K^4}\left|1+\epsilon^2\left(\alpha {\rm e}^{{\rm i}\phi_{1}}+\sum_{n=2}^{\frac{N+1}{2}}(A_{n}^2+B_{n}^2){\rm e}^{{\rm i}\phi_{n}}\right)\right|^2,
\end{equation}
where $\alpha$ is the same as in \eq~(\ref{eq:PT1E}). For the cases
$T=-1$, $A_{n}$ and $B_{n}$ are given by \eq~(\ref{eq:A_n}) and the
phases $\phi_{n}$ are given by \eq~(\ref{eq:phases}), whereas for the
cases $T=1$ we have
\begin{equation}\label{eq:A_nT1}
A_{n}=\frac{m_{\nu}}{u\sqrt{8N}\left[\sin^2\frac{(n-1)\pi}{N}-\frac{\lambda}{4}\right]}\quad{\rm
and}\quad
B_{n}=\frac{\sin\frac{(n-1)\pi}{N}}{\sqrt{2N}\left[\sin^2\frac{(n-1)\pi}{N}-\frac{\lambda}{4}\right]}.
\end{equation}
In this case, the phases are given by \eq~(\ref{eq:phases}), but with
the masses [\cf~\eq~(\ref{eq:Wilsonspectra})]
\begin{equation}
m_{n}^2=4u^2\sin^2\frac{(n-1)\pi}{N}.
\end{equation}
In \Figs~\ref{fig:plot1}--\ref{fig:plot4}, we have illustrated the
different neutrino transition survival probabilities in vacuum as
functions of $L/E$ for the different cases $T=\pm 1$ and $N$ odd/even
for some specific choices of $N$. In \Figs~\ref{fig:plot1} and
\ref{fig:plot2}, we have given the transition probabilities from
\eqs~(\ref{eq:P2})--(\ref{eq:PT1O}), where for presentation purposes, we have chosen $1-P_{ff}$ on the ordinate. From the validity requirements in
\eq~(\ref{eq:constraint1}) we know that Rayleigh--Schr\"odinger
perturbation theory will break down at some point. In \Figs~\ref{fig:plot3} and \ref{fig:plot4}, we
have therefore presented the curves from numerical
calculations. Nevertheless, at least qualitatively, the neutrino transition
survival probabilities show similar patterns for the analytical and the numerical calculations.

In what follows, our choice of parameters would correspond in the
ADD scenario to a 4D Planck scale $M_{Pl}=3.4 \cdot 10^{18}\:{\rm GeV}$, a SM Higgs
doublet VEV $\langle H
\rangle=174\:{\rm GeV}$, a Yukawa coupling $y=1$ between the active and the
bulk neutrinos, and a compactification radius of
$R^{-1}=0.1\:{\rm eV}$. In \Figs~\ref{fig:plot1} and \ref{fig:plot2},
the associated fundamental scale would be $M_{*}=1\:{\rm TeV}$, which
gives $m_{D}\simeq 5\cdot 10^{-5}\:{\rm eV}$. In
\Figs~\ref{fig:plot3} and \ref{fig:plot4}, the corresponding
fundamental scale would be $M_{*}=50\:{\rm TeV}$, which gives $m_D\simeq 2.5\cdot 10^{-3}\:{\rm eV}$. We have distinguished the cases
$m_{\nu}=0$ and $m_{\nu}\neq 0$. In \Figs~\ref{fig:plot2}
and \ref{fig:plot4}, we have set $m_{\nu}=0.01\:{\rm eV}$, whereas
in the other figures, we have set $m_{\nu}=0$. We have also made a
comparison for the cases $T=1$ with the corresponding survival
probability in the case of a continuous large extra dimension.
 
The qualitative behavior of the curves can be understood by looking at
the expressions for the survival probabilities, \ie, \eqs~(\ref{eq:P2})--(\ref{eq:PT1O}). From these
expressions we observe that the dominant effect will be given by the
lowest lying modes. For practical purposes one can then average over the
higher modes, which means that the essential behavior will be determined by
only a few low modes. Let us first consider the case $m_{\nu}=0$. We
obtain for $T=-1$ and $N$ even the survival probability,
 when only the first mode is non-averaged, as $P_{ff}=2\epsilon^2B_{1}^2/K^4\cos\phi_{1}+{\rm const.}$, where the amplitude and the frequency are given by
$2\epsilon^2B_{1}^2/K^4$ and $\phi_{1}$, respectively. For the other cases we obtain similar results.

First, we observe that the frequencies are proportional to $1/R^2$, \ie, a smaller radius gives faster oscillations.
We also note that the frequencies differ for the cases $T=1$ and $T=-1$. This is due to the different mass eigenvalues that appear in the phases. Thus, we have for the case $T=-1$, when only the first mode is non-averaged, that the frequency is
proportional to $N^2\sin^2 \pi/2N$. However, for the case $T=1$ the frequency is proportional to $N^2\sin^2 \pi/N$, which is roughly
four times larger than for the case $T=-1$. This can be directly seen
in \Fig~\ref{fig:plot1}. The figures also show a dependency of the frequency on
$N$. This is because the frequency in for example the case $T=-1$ is proportional to
$N^2\sin^2\pi/(2N)$. This function grows rapidly for small $N$ and
converges quickly to a fixed value, $\pi^2/4$. A similar relation holds for the case $T=1$. This effect is best visible when $N\leq \mathcal{O}(10)$.

Second, the amplitude is proportional to $m_D^2R^2$ so that a change of these parameters significantly affects the amplitude. This can be seen by comparing for example \Figs~\ref{fig:plot1} and \ref{fig:plot3}.
There is also a difference in amplitude between the cases $T=1$ and $T=-1$. For the case $T=-1$ the amplitude is proportional to $B_1$, where $B_n$ for $T=-1$ is given in \eq~(\ref{eq:A_n}). However, for the case $T=1$ the amplitude is proportional to $B_2$ where $B_n$ for $T=1$ is given in \eq~(\ref{eq:A_nT1}). Since $B_1 (T=-1)>B_2 (T=1)$ the amplitude will be larger for $T=-1$.    

Note that the case $T=1$ and $N=5$ differs in a significant way from the other cases. The aperiodic behavior of this curve is due to the large interference effect between the two lowest modes, which together give the dominating behavior. This effect can be seen in the relation between the factors $B_2$ and $B_3$. As we increase $N$, the effect of $B_3$ will be suppressed in comparison with $B_2$, so that we for large $N$ obtain a sinusoidal-like behavior. For the case $T=1$ and $N=4$ there is no large interference effect between low-lying modes, since the sum in this case only includes one term. For the case $T=-1$ the corresponding ratio which gives the interference effect is $B_2/B_1$. This ratio is smaller than the ratio $B_3/B_2$ for the case $T=1$. Thus, for $T=-1$ we do not observe any significant distortion of the periodicity.     

Let us now consider the effect of a non-zero $m_{\nu}$. For $T=-1$ we
note that there will be an effect provided that $m_{\nu}R$ is large
enough. If $m_{\nu}^2 \gtrsim 1/R^2$, the frequency will be determined by
$m_{\nu}$. However, for the case we have considered, we have chosen
$m_{\nu}=0.01\:{\rm eV}$ and $R^{-1}=0.1\:{\rm eV}$, which means that the frequency will mainly be determined by $1/R^2$. Thus, for the case $T=-1$ there will be no drastic changes, which can be seen when comparing
the upper rows of \Figs~\ref{fig:plot1} and \ref{fig:plot2}. For $T=1$, on the other hand, there will be a significant effect from $m_\nu$.
This is obvious from \eqs~(\ref{eq:PT1E}) and
(\ref{eq:PT1O}), where the survival probability expressions contain a
term $\alpha e^{i\phi_1}$, in which $\alpha$ is approximately given by
$N/(4\pi^2R^2m_{\nu}^2)$. If $m_{\nu}R$ is sufficiently small, then
this term will be the dominating term. Thus, the survival probability
will be proportional to $\cos (m_{\nu}^2 L/2E)$. This is
the case in the lower row of \Fig~\ref{fig:plot2}. Note that this effect is due to the presence of a zero mode, which is absent for $T=-1$.

If $N$ is increased, then the curves obtains a more jagged shape because of the interference of a large number of KK modes with different frequencies \cite{Dienes:1998sb}. However, essential properties such as the amplitude and the frequency quickly
stabilizes. Finally, we observe in \Figs~\ref{fig:plot1} and \ref{fig:plot3} that the case $T=1$
reproduces the continuum case \cite{Dvali:1999cn, Mohapatra:2001} as expected.

We have seen that one could, at least in principle, probe $T=\pm 1$
as well as the number of lattice sites $N$ through neutrino
oscillation experiments. For low $N$ the best probe of $N$ is through
the frequency.

\section{Summary and Conclusions}
\label{sec:disc}
In this paper, we have considered a model for neutrino oscillations in a
deconstructed $U(1)$ gauge theory defined on the boundary of a
two-dimensional disk. If the masses of the link fields connecting the center
with the boundary are of the order $\sim 1\:{\rm TeV}$ ($10^2\:{\rm TeV}$),
while the link fields on the boundary have masses of the order
$\sim 10^{15}\:{\rm GeV}$ ($10^{12}\:{\rm GeV}$), then the model
generates sub-mm lattice spacings between the sites on the boundary. This allows to obtain dynamically a non-gravitational large extra
dimension with only a few sites. Here, we have motivated the general
significance of large (sub-mm) lattice spacings by reviewing the strong coupling behavior of
gravity in local theory spaces for the example of two discrete gravitational
extra dimensions.

We have analyzed the mass and mixing properties of a latticized right-handed
neutrino, which propagates on the boundary of the disk and may be twisted or untwisted. A discrete cyclic symmetry, which can be gauged, allows to
treat the latticized right-handed neutrino as a Wilson fermion with vanishing
bare mass. At the same time, the cyclic symmetry also introduces in the
non-local theory space a local Yukawa interaction between the active SM
neutrinos and the latticized right-handed neutrino. As a consequence, the
model simulates key features of the 5D continuum theory for neutrinos in the
ADD scenario.

We have studied the neutrino oscillation effects of the latticized right-handed neutrino in terms of the survival probability $P_{ff}$ of a single active
flavor $f$ for the cases of a twisted (untwisted) lattice fermion and an
even (odd) number of sites on the boundary. By taking the continuum limit, we could exactly
reproduce known oscillation patterns of existing 5D continuum theory models.
The most direct probe of our model parameters is through the frequency of
$P_{ff}$. For example, if the number of lattice sites $N$ is small, then 
$P_{ff}$ can exhibit a strongly aperiodic behavior for odd $N$. Possible
``odd-even artifacts'', however, quickly disappear when $N$ becomes
large. Generally, twisted and untwisted field configurations can be
distinguished through the different associated frequencies of $P_{ff}$, which
is for an untwisted neutrino roughly four times larger than for a twisted one. The presence or absence of an active Majorana neutrino mass also affects
the oscillation patterns of twisted and untwisted neutrinos in qualitatively different ways. Generally, it should be noted, however, that in more elaborate models one would have to include three flavors (as well as matter effects).
It could also be necessary to take into account additional large extra
dimensions. Therefore, the results obtained from our model should be viewed
as comparatively qualitative.

The neutrino oscillation effects that are introduced by the KK neutrinos
could, in principle, be observed in present and future precision neutrino
oscillation experiments, such as for example KamLAND \cite{KamLAND},
Borexino \cite{Alimonti:2000xc}, or the proposed 
Double-CHOOZ \cite{Ardellier:2004ui} experiment. Borexino would be capable to search for new solar neutrino oscillation effects in an energy range
$E\lesssim 1\:{\rm MeV}$ not covered by Super-Kamiokande or SNO \cite{solar}. Our model could be tested at short baselines by (future)
$\overline{\nu}_e$ (or $\nu_e$) disappearance experiments with sensitivities
for mixing angles $\lesssim 0.2$ between the active and the KK-neutrinos.
Here, it could prove useful to employ also two-reactor-two-detector-setups
\cite{Huber:2004bh}, where one may perform measurements practically free from
the typical systematic uncertainties in the reactor neutrino fluxes. More generally, one can consider any
experiment, which probes the effect of sterile neutrinos, provided that
one can identify the masses and mixings properly.

The non-zero mixing between the SM Higgs $H$ and the scalar link and
site variables will lead to invisible decays $H\rightarrow W'W'$ (if these
processes are kinematically allowed), which can be checked at the LHC or a future linear collider.  

It is clear, that standard Big Bang nucleosynthesis \cite{wal} will be
affected by the presence of the KK neutrinos. However, the bounds from measurements
of the $^4{\rm He}$ abundance can be alleviated by
assuming a primordial lepton asymmetry \cite{foot} or with low
reheating temperature \cite{Gelmini:2004ah}. The constraints on the effective
number of neutrino species from large scale structure data in conjunction
with cosmic microwave background measurements \cite{el} may also be evaded by
such a lepton asymmetry \cite{Hannestad:2003ye}. Also note that, in this paper, we have assumed the
same constraints that apply to continuous large gravitational extra dimensions,
but it has been argued \cite{Arkani-Hamed:2001ca} that several of the
standard constraints could be relaxed for non-gravitational deconstructed
dimensions.

\section*{Acknowledgments}

We would like to thank K.S.~Babu, T.~Enkhbat, I. Gogoladze, and
M.D. Schwartz for useful comments and discussions. One of us (G.S.)
would like to thank the division of mathematical physics at the Royal
Institute of Technology (KTH), Stockholm, Sweden, for the warm
hospitality during the stays at KTH, where part of this work was
developed. This work was supported by the Swedish Research Council
(Vetenskapsr{\aa}det), Contract Nos.~621-2001-1611, 621-2002-3577
(T.O.), the G{\"o}ran Gustafsson Foundation (T.H. and T.O.), the
Magnus Bergvall Foundation (T.O. and G.S.), and the U.S. Department of
Energy under Grant Nos. DE-FG02-04ER46140 and DE-FG02-04ER41306
(G.S.).

\begin{figure}
\begin{center}
\includegraphics*[width=17 cm]{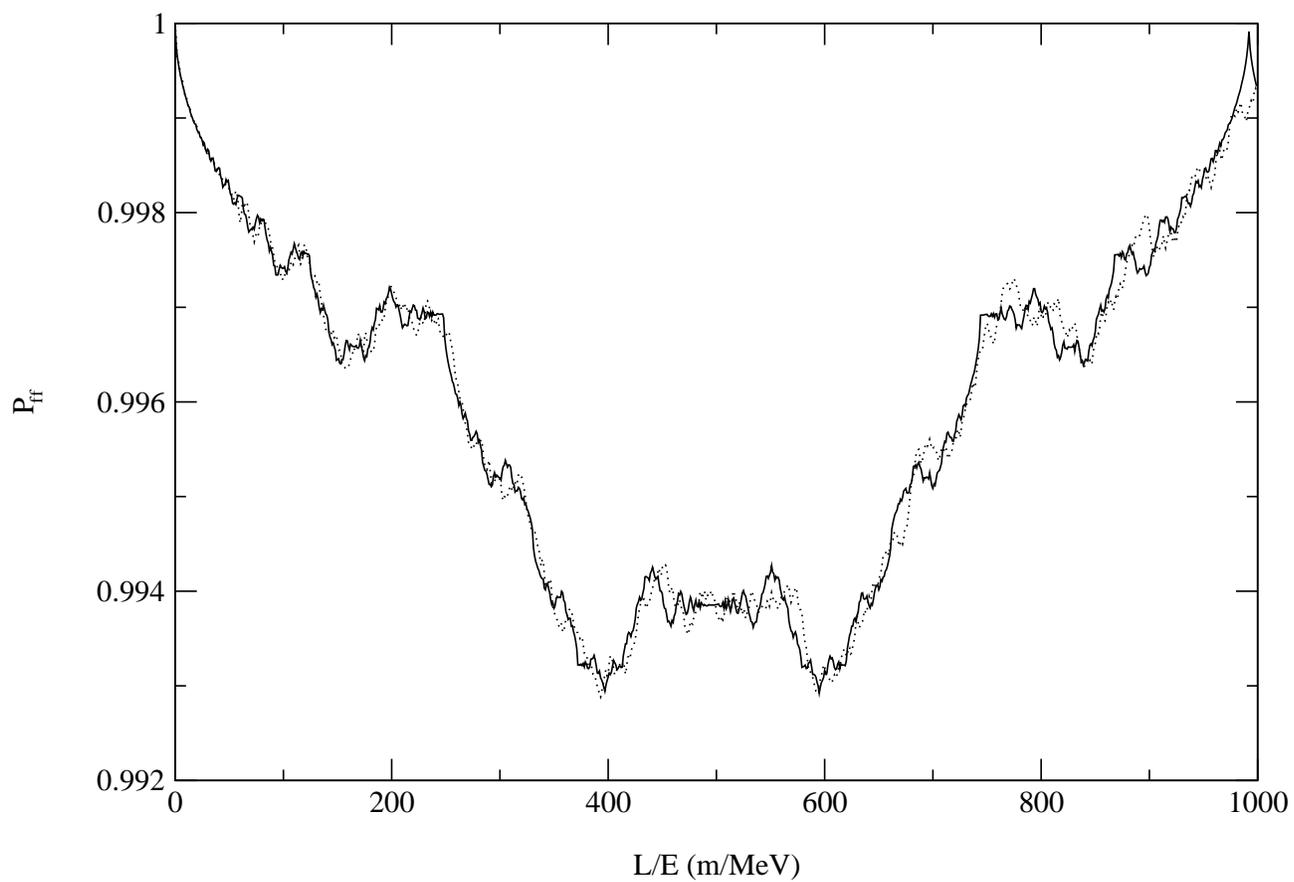}
\end{center}
\caption{\small{Amplification of the lower left panel in \Fig~\ref{fig:plot3}, comparing the continuum model (solid curve) with $N$=55 and $T=1$ (dotted curve)}}
\label{fig:bulkcont}
\end{figure}

\begin{appendix}

\section{Cancellation of anomalies}\label{app:anomalies}

The discrete $Z_{6M}$ symmetry in \eq~(\ref{eq:Z6M}) introduces on
each site on the boundary of the disk a parity-asymmetry between
$\nu_{nR}$ and $\nu_{nR}^c$. The model in
\Sec~\ref{sec:neutrinomasses} for the Wilson fermion is therefore
chiral. If we wish to gauge the $Z_{6M}$ symmetry, we will have to
ensure that the model remains free from chiral anomalies and that all
anomalous contributions from triangle diagrams to the
three-gauge-boson vertex functions cancel. In the low-energy effective
theory, the apparent unbroken $U(1)$ gauge symmetry\footnote{We neglect here the symmetry breaking introduced by the scalar site variables on the center (see \Sec~\ref{sec:Model}).} of the
deconstructed model would be $U(1)_{\rm diag}$ and it is only at short
distance scales that we become aware of the underlying enlarged
$U(1)^{N+1}$ gauge symmetry. It is interesting to compare in these two
limiting cases the formal cancellation of anomalous diagrams by
defining the gauge symmetry (containing the discrete factor) of our
model to be either
\begin{equation}\label{eq:models}
 {\rm model\:(a):}\quad G_{SM}\times U(1)_{\rm diag}\times
 Z_{6M}\quad{\rm or}\quad{\rm model\:(b):}\quad G_{SM}\times
 U(1)^{N+1}\times Z_{6M}.
\end{equation}
Implicitly, model (a) becomes equivalent with a non-linear sigma model
approximation of the deconstructed model in \Sec~\ref{sec:Model}. Of
course, if we are interested in the UV completing linear sigma model
description, only the anomalies calculated in model (b) are of
relevance.

\begin{figure}
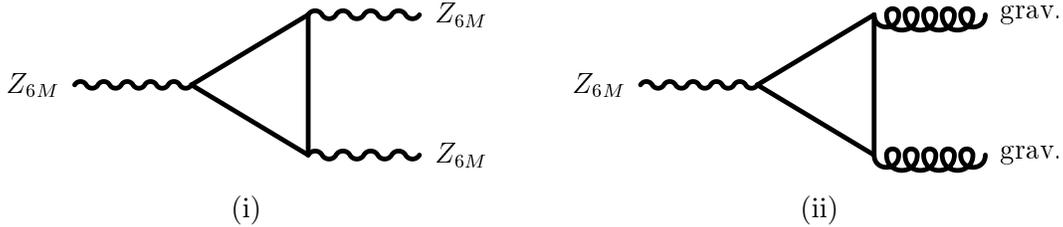

 \begin{center}
 \begin{tabular}{ccc}
 \includegraphics*[bb= 74 363 275 439,height=2.5cm]{figures.ps}&
 \hspace*{1mm}&
 \includegraphics*[bb= 333 363 538 439,height=2.5cm]{figures.ps}\\
 \small{(i)}&&\small{(ii)}
\end{tabular} 
\vspace*{-2mm}
\caption{\small{Triangle diagrams leading to cubic (i) and
gauge-gravitational (ii) anomalies in the models (a) and (b). These
anomalies cancel between neighboring sites.}}\label{fig:cubictrace}
\end{center}
\end{figure}

To work out the anomaly cancellation for models (a) and (b) in
\eq~(\ref{eq:models}), we will denote the $U(1)_n$, $U(1)_{\rm diag}$,
and $Z_{6M}$ charges of a field $f$ by $q_i(f)$, $q_{\rm diag}(f)$,
and $q_{Z_{6M}}(f)$, respectively. First, we observe that the fermions
$f$ located on the center of the disk -- \ie, the SM fermions
and the fields $N_\alpha$ -- satisfy $q_{0}(f)=q_{\rm
diag}(f)=q_{Z_{6M}}(f)$. Thus, the SM fermions and the $N_\alpha$ do
not contribute to any anomalies and we can from now on concentrate on the
anomalous diagrams involving only the right-handed neutrinos
$\nu_{Rn}$ and $\overline{\nu^c_R}_n$.  Since the $U(1)_{\rm diag}$
and $U(1)^{N+1}$ gauge bosons couple equally to $\nu_{Rn}$ and
$\overline{\nu_R^c}_n$ on each site on the boundary, all anomalies
which do not involve a $Z_{6M}$ coupling vanish
automatically. Consider now the triangle diagrams in
\Fig~\ref{fig:cubictrace}, which do not have any $U(1)_i$ or
$U(1)_{\rm diag}$ gauge bosons at their vertices.
If we choose in Eq.~(\ref{eq:Z6M}) $M=(N+2)^2-4$, then in both models (a) and (b), the cubic $[Z_{6M}]^3$ anomaly (i)
and the gauge-gravitational anomaly (ii) are proportional to
\begin{equation}\label{eq:cubictrace}
 \sum_{n=1}^N[q_{Z_{6M}}(\nu_{Rn})^3+q_{Z_{6M}}(\nu_{Rn}^c)^3]=0\quad{\rm and}
\quad\sum_{n=1}^N[q_{Z_{6M}}(\nu_{Rn})+q_{Z_{6M}}(\nu_{Rn}^c)]=0,
\end{equation}
where we have used that $q_{Z_{6M}}(\nu_{R(n-1)})\equiv
-q_{Z_{6M}}(\nu_{Rn}^c)$ and $q_{Z_{6M}}(\nu_{R0})\equiv
q_{Z_{6M}}(\nu_{RN})$, implying that the anomalies cancel between
neighboring sites and vanish when summing over all $2N$ right-handed
neutrino species.

\begin{figure}
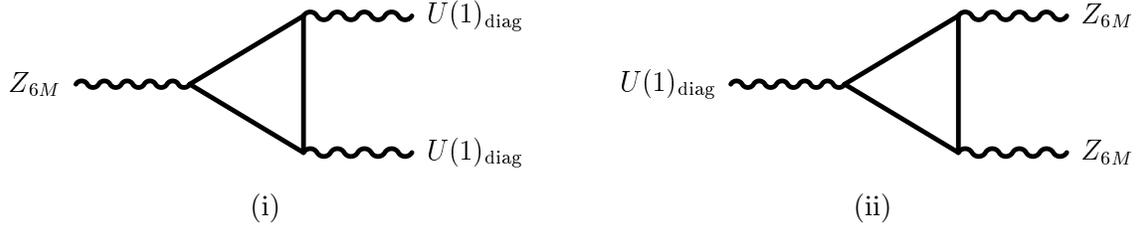

 \begin{center}
 \begin{tabular}{ccc}
 \includegraphics*[bb= 74 248 297 326,height=2.5cm]{figures.ps}&
 \hspace*{1mm}&
 \includegraphics*[bb= 313 248 535 326,height=2.5cm]{figures.ps}\\
 \small{(i)}&&\small{(ii)}
\end{tabular} 
\vspace*{-2mm}
 \caption{\small{Cancellation of mixed anomalies in model (a). The
$Z_{6M}[U(1)_{\rm diag}]^2$ anomaly (i) and the
$U(1)_{\rm diag}[Z_{6M}]^2$ anomaly (ii) vanish when summing over all
sites.}}\label{fig:mixedanomalies}
\end{center}
\end{figure}

The anomaly cancellations differ substantially between model (a) and
(b) when evaluating the triangle diagrams of the type shown in
\Fig~\ref{fig:mixedanomalies}, which have at least one $U(1)$ gauge
boson at one of their vertices.
Let us first restrict to model (a). In \Fig~\ref{fig:mixedanomalies},
the mixed $Z_{6M}[U(1)_{\rm diag}]^2$ anomaly (i) and the
$U(1)_{\rm diag}[Z_{6M}]^2$ anomaly (ii) are proportional to
\begin{equation}
 \sum_{n=1}^N[q_{Z_{6M}}(\nu_{Rn})+q_{Z_{6M}}(\nu_{Rn}^c)]=0\quad{\rm and}
\quad\sum_{n=1}^N[q_{Z_{6M}}(\nu_{Rn})^2-q_{Z_{6M}}(\nu_{Rn}^c)^2]=0,
\end{equation}
where we have used that $q_{\rm diag}(\nu_{Rn})\equiv -q_{\rm
diag}(\nu_{Rn}^c)=-1$. Again, the diagrams cancel between neighboring
sites and vanish when summing over all sites. In total, we thus find
that all divergent triangle diagrams in model (a) formally add up to
zero. Let us next consider the corresponding anomalies in model (b).
The dangerous mixed $Z_{6M}[U(1)_n]^2$ and $U(1)_n[Z_{6M}]^2$
anomalies are obtained from the diagrams in
\Fig~\ref{fig:mixedanomalies} by replacing in (i) and (ii) the gauge
bosons according to $U(1)_{\rm diag}\rightarrow U(1)_n$. In this case,
the diagrams (i) and (ii) in \Fig~\ref{fig:mixedanomalies} become
divergent and the summation over all sites does not remove the
divergences, since the diagrams belonging to different lattice sites
have different external legs and are thus inequivalent.

In order to remove the anomalies in model (b), we add on each site of
the boundary extra fermions with appropriate quantum numbers. We place
on the site corresponding to the gauge group $U(1)_n$
($n=1,2,\ldots,N$) three additional Dirac spinors, which are written
component-wise in the Weyl basis as
$\tilde{\Psi}_{n}\equiv(\tilde{\nu}_{n},
\overline{\tilde{\nu}}_n^c)^T$,
$X_n\equiv(\eta_n,\overline{\eta}_n^c)^T$, and $\tilde{X}_n\equiv
(\tilde{\eta}_n,\overline{\tilde{\eta}}_n^c)^T$. The fields
$\tilde{\Psi}$, $X_n$, and $\tilde{X}_n$ carry the $U(1)_n$ charges
$+1$, $-(n+1)$, and $n+2$, respectively, and are singlets under
$G_{SM}$ and the other gauge groups $U(1)_{i\neq n}$. In addition, we
assume that the extra fermions carry specific $Z_{6M}$ charges. The
$U(1)_n$ and $Z_{6M}$ charge assignment for all SM singlet fermions is
summarized in \Tab~\ref{tab:charges}.
\begin{table}
\begin{center}
\begin{tabular}{|c|c|c|c|c|c|c|c|c|}
 \hline 
 field & $\nu_{Rn}$ & $\nu_{Rn}^c$ & $\tilde{\nu}_n$ &
 $\tilde{\nu}_n^c$ & $\eta_n$ & $\eta_n^c$ & $\tilde{\eta}_n$ &
$\tilde{\eta}_n^c$\\
\hline
\hline
 $q_n$ & $-1$ & $+1$ & $+1$ & $-1$ & $-(n+1)$ & $n+1$ & $n+2$ & $-(n+2)$\\
\hline
$q_{Z_{6M}}$ & $(n+2)^2$ & $-(n+1)^2$ & $(n+2)^2$ & $-(n+1)^2$&$+1$&$+1$&
$-1$ & $-1$\\
\hline
\end{tabular}
\caption{\small $U(1)_n$ $(n=1,2,\ldots,N)$ and $Z_{6M}$ charges of
the fermions on the boundary of the disk. These fields transform
trivially under the other gauge groups $U(1)_{i\neq n}$. The $Z_{6M}$
charges have been normalized with respect to the SM leptons in
\eq~(\ref{eq:Z6M}).}
\label{tab:charges}
\end{center}
\end{table}

>From \Tab~\ref{tab:charges}, we find that the cubic and
gauge-gravitational $U(1)_n$ and $Z_{6M}$ anomalies still vanish, since the
$U(1)_n$ and $Z_{6M}$ symmetries satisfy for the extra fermions relations similar
to \eq~(\ref{eq:cubictrace}). The mixed $Z_{6M}[U(1)_n]^2$ anomaly is
proportional to
\begin{eqnarray}\label{eq:ZU2}
\sum_{n=1}^N\left\{[q_{Z_{6M}}(\nu_{Rn})-q_n(\tilde{\eta}_n)^2]
+[q_{Z_{6M}}(\nu_{Rn}^c)+q_n(\eta_n^c)^2]\right.&&\nonumber\\
\left.+[q_{Z_{6M}}(\tilde{\nu}_n)-q_n(\tilde{\eta}_n^c)^2]
+[q_{Z_{6M}}(\tilde{\nu}_n^c)+q_n(\eta_n)^2]\right\}&=&0,
\end{eqnarray}
where each bracketed term inside the sum is zero. Therefore, these anomalies
cancel on each site.
The mixed $U(1)_n[Z_{6M}]^2$ anomaly is proportional to
\begin{eqnarray}\label{eq:UZ2}
\sum_{n=1}^N\left\{[-q_{Z_{6M}}(\nu_{Rn})^2+q_{Z_{6M}}(\tilde{\nu}_n)^2]
+[q_n(\eta_n)+q_n(\eta_n^c)]\right.&&\nonumber\\
\left.+
[q_{Z_{6M}}(\nu_{Rn}^c)^2-q_{Z_{6M}}(\tilde{\nu}_n^c)^2]
+[q_n(\tilde{\eta}_n)+q_n(\tilde{\eta}_n^c)]\right\}&=&0,
\end{eqnarray}
where each bracketed expression inside the sum vanishes. Again, all
anomalies cancel individually on each lattice site. In total, the
$Z_{6M}$ symmetry is therefore anomaly-free. In addition, this model
is chiral, since the $U(1)^{N+1}\times Z_{3M}$ symmetry forbids any
bare mass terms for the fermions.

\section{Neutrino mixing matrices}\label{app:diagonalization}

In the basis $(\nu,\nu_{R1},\nu_{R2},\ldots,\nu_{RN},\nu^c_{R1},
\nu^c_{R2},\ldots,\nu^c_{RN})$, the total neutrino mass matrix $M$
described by the action density $\mathcal{L}^\nu_{\rm m}$ in
\eq~(\ref{eq:mass+mixing}) takes the form
\begin{equation}\label{eq:M} 
M=\left(
\begin{matrix}
 M_L & M_D\\
 M_D^T & 0
\end{matrix}
\right),
\end{equation}
where $M_L$ and $M_D$ denote $(N+1)\times(N+1)$ and
$(N+1)\times N$ matrices respectively, which are explicitly given by
\begin{equation}\label{eq:M_L}
M_L=
\left(
\begin{matrix}
 m_\nu & \sqrt{N}m_D & 0 & \cdots & 0\\
 \sqrt{N}m_D & 0 & 0 & \cdots & 0 \\
 0  & 0 & 0 & \cdots & 0\\
 \vdots & \vdots & \vdots& \ddots &\vdots\\
 0 & 0 & 0 & \cdots & 0
\end{matrix}
\right),\quad
 M_D=u \left(
\begin{matrix}
 0 &  0 & 0 & \cdots   & 0\\
-1 & +1 & 0 &    & \\
0  & -1 & +1&    &  \\
   & \ddots   & \ddots & \ddots\\
   &    &    0    &  -1 & +1\\
T  &    &     &       0 &  -1
\end{matrix}
\right).
\end{equation}
Here, the Majorana-like matrix $M_L$ is defined in the basis
$(\nu,\nu_{R1},\nu_{R2},\ldots,\nu_{RN})$, whereas the Dirac-like
matrix $M_D$ is spanned by $(\nu,\nu_{R1},\nu_{R2},\ldots,\nu_{RN})$
and $(\nu^c_{R1}, \nu^c_{R2},\ldots,\nu^c_{RN})$. In \eq~(\ref{eq:M}),
``0'' denotes an $N\times N$ matrix with zero entries only. The
neutrino masses and mixing angles can be determined from the product
$MM^{\dagger}$, which in this basis explicitly reads
\begin{equation}\label{eq:MMdagger}
{\small MM^{\dagger}=u^2\left(
\begin{array}{cccccccccccc}
 \frac{m_{\nu}^2 + Nm_{D}^2}{u^2} & \frac{\sqrt{N}m_{\nu}m_{D}}{u^2} & 0 & 0 & \cdots
 & 0 & \vline & -\frac{\sqrt{N}m_{D}}{u} & \frac{\sqrt{N}m_{D}}{u} & 0 & \cdots & 0\\
 \frac{\sqrt{N}m_{\nu}m_{D}}{u^2} & 2 +\frac{Nm_{D}^2}{u^2} & -1 & & & -T &
 \vline & 0 & 0 & 0 & \cdots & 0\\
 0 & -1 & 2 & -1 & & & \vline & 0 & 0 & 0 & \cdots & 0\\
 \vdots & & \ddots & \ddots & \ddots & & \vline & \vdots & \vdots & \vdots &
 \ddots & \vdots\\
 0 & & & -1 & 2 & -1 & \vline & 0 & 0 & 0 & \cdots & 0\\ 
 0 & -T & & & -1 & 2 & \vline & 0 & 0 & 0 & \cdots & 0\\
 \hline & & & & & & \vline & & & & &\\
 -\frac{\sqrt{N}m_{D}}{u} & 0 & 0 & \cdots & 0 & 0 & \vline & 2 & -1 & & & -T\\
 \frac{\sqrt{N}m_{D}}{u} & 0 & 0 & \cdots & 0 & 0 & \vline & -1 & 2 & -1 & &\\
 0 & 0 & 0 & \cdots & 0 & 0 & \vline & & \ddots & \ddots & \ddots &\\
 \vdots & \vdots & \vdots & \ddots & \vdots & \vdots & \vline & & & -1
 & 2 & -1\\
 0 & 0 & 0 & \cdots & 0 & 0 & \vline & -T & & & -1 & 2\\
\end{array}
\right),}
\end{equation}
where $T = \pm 1$ and the blank entries are zero. Next, we want to diagonalize this matrix. Since
$m_\nu, \sqrt{N} m_D \ll u$, we can define the quantity
$\epsilon\equiv \sqrt{N}m_{D}/u \ll 1$, which we will use as an
expansion parameter in perturbation theory. We will diagonalize the
matrix $MM^{\dagger}$ in steps. First, we perform the transformation
$MM^{\dagger}\rightarrow U^TMM^\dagger U^\ast$ using the
block-diagonal $(2N+1)\times (2N+1)$ matrix $U\equiv{\rm
diag}(1,U_1,U_1)$, where $U_1$ denotes a unitary $N\times N$
matrix. For $T=-1$ and $N$ even, the matrix $U_1$ reads
\begin{equation}\label{eq:UU1}
{\footnotesize
U_1 = \sqrt{\frac{2}{N}} \left(\begin{matrix}
0 & 0 & \cdots & 0 & 1 & \cdots & 1\\
\sin\frac{\pi}{N} & \sin\frac{3\pi}{N} & \cdots & \sin\frac{(N-1)\pi}{N} &
\cos\frac{(N+1)\pi}{N} & \cdots & \cos\frac{(2N-1)\pi}{N}\\
\sin\frac{2\pi}{N} & \sin\frac{6\pi}{N} & \cdots & \sin\frac{2(N-1)\pi}{N} &
\cos\frac{2(N+1)\pi}{N} & \cdots & \cos\frac{2(2N-1)\pi}{N}\\
\vdots & \vdots & \ddots & \vdots & \vdots & \ddots & \vdots\\
\sin\frac{(N-1)\pi}{N} & \sin\frac{3(N-1)\pi}{N} & \cdots & \sin\frac{(N-1)(N-1)\pi}{N} &
\cos\frac{(N-1)(N+1)\pi}{N} & \cdots & \cos\frac{(N-1)(2N-1)\pi}{N}\\
\end{matrix}\right),}
\end{equation}
whereas for $T=-1$ and $N$ odd, we have
\begin{equation}
{\footnotesize
U_{1} = \sqrt{\frac{2}{N}} \left(\begin{matrix} 0 & 0 & \cdots & 0 & \frac{1}{\sqrt{2}} & 1 & \cdots & 1 \\
\sin\frac{\pi}{N} & \sin\frac{3\pi}{N} & \cdots & \sin\frac{(N-2)\pi}{N} & -\frac{1}{\sqrt{2}} & \cos\frac{(N+2)\pi}{N} & \cdots & \cos\frac{(2N-1)\pi}{N}\\
\sin\frac{2\pi}{N} & \sin\frac{6\pi}{N} & \cdots & \sin\frac{2(N-2)\pi}{N} & \frac{1}{\sqrt{2}} & \cos\frac{2(N+2)\pi}{N} & \cdots & \cos\frac{2(2N-1)\pi}{N} \\
\vdots & \vdots & \ddots & \vdots & \vdots & \vdots & \ddots & \vdots \\
\sin\frac{(N-1)\pi}{N} & \sin\frac{3(N-1)\pi}{N} & \cdots & \sin\frac{(N-1)(N-2)\pi}{N} & \frac{1}{\sqrt{2}} & \cos\frac{(N-1)(N+2)\pi}{N} & \cdots & \cos\frac{(N-1)(2N-1)}{N}\\
\end{matrix} \right) .}
\end{equation} 
Similarly, for $T=1$ and $N$ even, we have
\begin{equation}
{\footnotesize
U_{1} = \sqrt{\frac{2}{N}} \left(\begin{matrix}
\frac{1}{\sqrt{2}} & 0 & \cdots & 0 & \frac{1}{\sqrt{2}} & 1 & \cdots
& 1\\ \frac{1}{\sqrt{2}} & \sin\frac{2\pi}{N} & \cdots &
\sin\frac{(N-2)\pi}{N} & -\frac{1}{\sqrt{2}} & \cos\frac{(N+2)\pi}{N}
& \cdots & \cos\frac{(2N-2)\pi}{N}\\ \frac{1}{\sqrt{2}} &
\sin\frac{4\pi}{N} & \cdots & \sin\frac{2(N-2)\pi}{N} &
\frac{1}{\sqrt{2}} & \cos\frac{2(N+2)\pi}{N} & \cdots &
\cos\frac{2(2N-2)\pi}{N} \\ \vdots & \vdots & \ddots &\vdots &\vdots &
\vdots & \ddots & \vdots \\ \frac{1}{\sqrt{2}} &
\sin\frac{2(N-1)\pi}{N} & \cdots & \sin\frac{(N-1)(N-2)\pi}{N} &
-\frac{1}{\sqrt{2}} & \cos\frac{(N-1)(N+2)\pi}{N} & \cdots &
\cos\frac{(N-1)(2N-2)\pi}{N}\\
\end{matrix}\right),}
\end{equation}
and finally, for $T=1$ and $N$ odd, we have
\begin{equation}
{\footnotesize
U_{1} = \sqrt{\frac{2}{N}} \left(\begin{matrix} \frac{1}{\sqrt{2}} & 0 & \cdots & 0 & 1 & \cdots & 1 \\
\frac{1}{\sqrt{2}} & \sin\frac{2\pi}{N}  & \cdots & \sin\frac{(N-1)\pi}{N} & \cos\frac{(N+1)\pi}{N} & \cdots & \cos\frac{(2N-2)\pi}{N} \\
\frac{1}{\sqrt{2}} & \sin\frac{4\pi}{N}& \cdots & \sin\frac{2(N-1)\pi}{N} & \cos\frac{2(N+1)\pi}{N} & \cdots & \cos\frac{2(2N-2)\pi}{N}\\
\vdots & \vdots & \ddots & \vdots &  \vdots & \ddots & \vdots \\
\frac{1}{\sqrt{2}} & \sin\frac{2(N-1)\pi}{N}  & \cdots & \sin\frac{(N-1)(N-1)\pi}{N} & \cos\frac{(N-1)(N+1)\pi}{N} & \cdots & \cos\frac{(N-1)(2N-2)\pi}{N}\\
\end{matrix}\right) . }
\end{equation}
In \Sec~\ref{sec:bulkmodes}, we consider the case $T=-1$ and $N$
even. The other cases follow in similar ways. The rotation from the
matrix $U_1$ in \eq~(\ref{eq:UU1}) has the effect of diagonalizing
the ``gauge-boson-like'' submatrices in \eq~(\ref{eq:MMdagger}) and
leads in the new basis to the matrix in \eq~(\ref{eq:X}), which can
then be further diagonalized using perturbation theory as described in
\Sec~\ref{sec:bulkmodes}.

\end{appendix}

\end{document}